%% file: NIKA_tSZ_observation.tex
\documentclass[twocolumn,structabstract]{aa}  
\usepackage{fixltx2e}
\usepackage[english]{babel}
\usepackage{graphicx,amsmath}
\usepackage{epstopdf}
\usepackage{epsf,color}
\usepackage[mathscr]{eucal}
\usepackage{amsmath}
\usepackage{amssymb,amsfonts}
\usepackage{natbib}
\usepackage{graphicx}
\usepackage{txfonts}
\usepackage{dsfont}
\usepackage[breaklinks, colorlinks, citecolor=blue, linkcolor=MyBlue, urlcolor=RoyalPurple, colorlinks=true, linkcolor=blue, debug, baseurl=' ']{hyperref}

\def\simlt{\lower.5ex\hbox{$\; \buildrel < \over \sim \;$}}
\def\simgt{\lower.5ex\hbox{$\; \buildrel > \over \sim \;$}}

\bibpunct{(}{)}{;}{a}{}{,} 
\begin{document}
\title{First observation of the thermal Sunyaev-Zel'dovich effect with kinetic inductance detectors}
\input{listeauthorsSZ}
\date{Received \today \ / Accepted --}
\input{00_abstract}
\titlerunning{First tSZ observation with KIDs}
\authorrunning{R.~Adam, B.~Comis, J.~F.~Mac\'ias-P\'erez, et al.}
\maketitle
\section{Introduction}
\label{sec:introduction}
\input{01_introduction}
\section{Previous observations of RX~J1347.5-1145}
\label{sec:previous_obs}
\input{02_previous_obs.tex}
\section{Observations with {\itshape \bfseries NIKA}}
\label{sec:observations}
\input{03_nika_obs.tex}
\section{Thermal Sunyaev-Zel'dovich dedicated data analysis and mapmaking}
\label{sec:SZ_analysis}
\input{04_sz_ana.tex}

\section{Validation of the analysis}
\label{sec:simu_and_idcs}
\input{05_simu.tex}

\section{Results}
\label{sec:results}
\input{06_result.tex}
\section{Comparison to other external data sets}
\label{sec:comparison}
\input{07_comparison.tex}

\section{Conclusions and prospectives for {\itshape \bfseries NIKA2}}
\label{sec:conclusion}
\input{08_conclusion.tex}
\begin{acknowledgements}
We would like to thank the IRAM staff for their support during the campaign. 
This work has been partially funded by the Foundation Nanoscience Grenoble, the ANR under the contracts "MKIDS" and "NIKA". 
This work has been partially supported by the LabEx FOCUS ANR-11-LABX-0013. 
This work has benefited from the support of the European Research Council Advanced Grant ORISTARS under the European Union's Seventh Framework Programme (Grant Agreement no. 291294).
The NIKA dilution cryostat has been designed and built at the Institut N\'eel. In particular, we acknowledge the crucial contribution of the Cryogenics Group, and in particular Gregory Garde, Henri Rodenas, Jean Paul Leggeri, Philippe Camus. 
R. A. would like to thank the ENIGMASS French LabEx for funding this work. 
B. C. acknowledges support from the CNES post-doctoral fellowship program. 
E. P. acknowledges the support of grant ANR-11-BS56-015. 
We gratefully thank the anonymous referee for useful comments that have improved not only the quality of the paper but that will also help future analysis of {\it NIKA} tSZ observations. 
Finally we would like to tank C\'eline Combet for a careful reading of the paper.
\end{acknowledgements}
\bibliography{biblio}
\end{document}

%% file: listeauthorsSZ.tex
\author{R.~Adam\inst{\ref{inst1}}
\and B.~Comis\inst{\ref{inst1}}
\and J.~F.~Mac\'ias-P\'erez\inst{\ref{inst1}}
\and A.~Adane\inst{\ref{inst2}}
\and P.~Ade\inst{\ref{inst3}}
\and P.~Andr\'e\inst{\ref{inst4}}
\and A.~Beelen\inst{\ref{inst5}}
\and B.~Belier\inst{\ref{inst6}}
\and A.~Beno\^it\inst{\ref{inst7}}
\and A.~Bideaud\inst{\ref{inst3}}
\and N.~Billot\inst{\ref{inst8}}
\and N.~Boudou\inst{\ref{inst7}}
\and O.~Bourrion\inst{\ref{inst1}}
\and M.~Calvo\inst{\ref{inst7}}
\and A.~Catalano\inst{\ref{inst1}}
\and G.~Coiffard\inst{\ref{inst2}}
\and A.~D'Addabbo\inst{\ref{inst7}, \ref{inst14}}
\and F.-X.~D\'esert\inst{\ref{inst9}}
\and S.~Doyle\inst{\ref{inst3}}
\and J.~Goupy\inst{\ref{inst7}}
\and C.~Kramer\inst{\ref{inst8}}
\and S.~Leclercq\inst{\ref{inst2}}
\and J.~Martino\inst{\ref{inst5}}
\and P.~Mauskopf\inst{\ref{inst3}, \ref{inst13}}
\and F.~Mayet\inst{\ref{inst1}}
\and A.~Monfardini\inst{\ref{inst7}}
\and F.~Pajot\inst{\ref{inst5}}
\and E.~Pascale\inst{\ref{inst3}}
\and L. Perotto\inst{\ref{inst1}}
\and E.~Pointecouteau\inst{\ref{inst10}, \ref{inst11}}
\and N.~Ponthieu\inst{\ref{inst9}}
\and V.~Rev\'eret\inst{\ref{inst4}}
\and L.~Rodriguez\inst{\ref{inst4}}
\and G.~Savini\inst{\ref{inst12}}
\and K.~Schuster\inst{\ref{inst2}}
\and A.~Sievers\inst{\ref{inst8}}
\and C.~Tucker\inst{\ref{inst3}}
\and R.~Zylka\inst{\ref{inst2}}}

\offprints{R. Adam - adam@lpsc.in2p3.fr}

\institute{
Laboratoire de Physique Subatomique et de Cosmologie,
Universit\'e Joseph Fourier Grenoble 1,
  CNRS/IN2P3, Institut Polytechnique de Grenoble, 
  53, rue des Martyrs, Grenoble, France
  \label{inst1}
\and
Institut de RadioAstronomie Millim\'etrique (IRAM), Grenoble, France
  \label{inst2}
\and
Astronomy Instrumentation Group, University of Cardiff, UK
  \label{inst3}
\and
Laboratoire AIM, CEA/IRFU, CNRS/INSU, Universit\'e Paris Diderot, CEA-Saclay, 91191 Gif-Sur-Yvette, France 
  \label{inst4}
\and
Institut d'Astrophysique Spatiale (IAS), CNRS and Universit\'e Paris Sud, Orsay, France
  \label{inst5}
\and
Institut d'Electronique Fondamentale (IEF), Universit\'e Paris Sud, Orsay, France
  \label{inst6}
\and
Institut N\'eel, CNRS and Universit\'e de Grenoble, France
  \label{inst7}
\and
Institut de RadioAstronomie Millim\'etrique (IRAM), Granada, Spain
  \label{inst8}
\and
Institut de Plan\'etologie et d'Astrophysique de Grenoble (IPAG), CNRS and Universit\'e de
Grenoble, France
  \label{inst9}
\and
Universit\'e de Toulouse, UPS-OMP, Institut de Recherche en Astrophysique et Plan\'etologie (IRAP), Toulouse, France
  \label{inst10}
\and
CNRS, IRAP, 9 Av. colonel Roche, BP 44346, F-31028 Toulouse cedex 4, France 
  \label{inst11}
\and
University College London, Department 
of Physics and Astronomy, Gower Street, London WC1E 6BT, UK
  \label{inst12}
  \and
  School of Earth and Space Exploration and Department of Physics, 
	Arizona State University, Tempe, AZ 85287
  \label{inst13}
    \and
  Dipartimento di Fisica, Sapienza Universit\`a di Roma, Piazzale Aldo Moro 5, I-00185 Roma, Italy
  \label{inst14}
}

%% file: 00_abstract.tex
\abstract {Clusters of galaxies provide valuable information on the evolution of the Universe and large scale structures. Recent cluster observations via the thermal Sunyaev-Zel'dovich (tSZ) effect have proven to be a powerful tool to detect and study them. In this context, high resolution tSZ observations ($\sim$ tens of arcsec) are of particular interest to probe intermediate and high redshift clusters.}
{Observations of the tSZ effect will be carried out with the millimeter dual-band {\it NIKA2 }camera, based on Kinetic Inductance Detectors (KIDs) to be installed at the IRAM 30-meter telescope in 2015. To demonstrate the potential of such an instrument, we present tSZ observations with the {\it NIKA }camera prototype, consisting of two arrays of 132 and 224 detectors that observe at 140 and 240~GHz with a 18.5 and 12.5~arcsec angular resolution, respectively.} 
{The cluster \mbox{RX~J1347.5-1145} was observed simultaneously at 140 and 240~GHz. We used a spectral decorrelation technique to remove the atmospheric noise and obtain a map of the cluster at 140~GHz. The efficiency of this procedure has been characterized through realistic simulations of the observations.}
{The observed 140~GHz map presents a decrement at the cluster position consistent with the tSZ nature of the signal. We used this map to study the pressure distribution of the cluster by fitting a gNFW model to the data. Subtracting this model from the map, we confirm that \mbox{RX~J1347.5-1145} is an ongoing merger, which confirms and complements previous tSZ and X-ray observations.}
{For the first time, we demonstrate the tSZ capability of KID based instruments. The {\it NIKA2 }camera with $\sim 5000$ detectors and a $6.5$~arcmin field of view will be well-suited for in-depth studies of the intra cluster medium in intermediate to high redshifts, which enables the characterization of  recently detected clusters by the {\it Planck }satellite.}

%% file: 01_introduction.tex
Galaxy clusters are the largest gravitationally bound objects in the Universe. Their formation strongly depends on the content and the history of the Universe within the framework of a bottom-up scenario \citep[{\it e.g.,} ][]{kravtsov_2012}, where there is merging of small clusters to form larger ones. They are classically probed using \mbox{X-ray} produced via bremsstrahlung emission of the electrons in the intracluster medium (ICM) but are also measured in the optical and infrared wavelengths, which trace the stellar populations in the member galaxies. Their radio emission is related to the acceleration of charged particles, and the lensing of background objects provides surface mass density measurements from multi-band optical and infrared data. See \cite{bohringer_2010, gal_2006, oliver_2012, feretti_2012, kneib_2011} for reviews on the different cluster observables.

The thermal Sunyaev-Zel'dovich (tSZ) effect \citep{SunyaevZeldovich1, SunyaevZeldovich2}, which consists of the inverse Compton scatter of Cosmic Microwave Background (CMB) photons on hot electrons in the ICM, can be used as a complementary method to probe galaxy clusters \citep[see][for a detailed review on the tSZ effect]{Birkinshaw, Carlstrom_et_al_2002}. Three-dimensional information on the cluster may be inferred using the characteristic dependences of \mbox{X-ray} (sensitive to the line-of-sight integral of the density squared and the square root of the temperature) and tSZ (sensitive to the integrated pressure along the line-of-sight) with the properties of the ICM. This gives a more accurate picture than \mbox{X-ray} or tSZ alone, especially in the case of merging systems \citep{Basu}. In addition, unlike other observational approaches, the tSZ signal is not affected by cosmological dimming. Only the angular size of the observed cluster depends on the distance to the source. High angular resolution tSZ observations are therefore of particular interest to probe structure formation at high redshift.

The resolutions of the main current instruments measuring the tSZ effect are of the order of the~arcmin. It is larger than 5~arcmin for the {\it Planck} satellite \citep{PLANCK_mission} and about 1~arcmin for the South Pole Telescope \citep[{\it SPT};][]{SPT} and the Atacama Cosmology Telescope \citep[{\it ACT};][]{ACT}. Higher resolution instruments, such as {\it MUSTANG} \citep[$\sim 8$~arcsec resolution at 90~GHz;][]{mason_2010, korngut_2011}, may suffer from filtering of large-scale structures due to the atmospheric noise removal when observing at a single frequency band. High redshift tSZ observations, therefore, need a new generation of instruments. The New IRAM KID Arrays ({\it NIKA}) is a prototype of a high-resolution camera based on Kinetic Inductance Detectors (KIDs) \citep{day_2003,KID_for_CMB} in development for millimeter wave astronomy \citep{NIKA_2011}. It consists of two arrays of 132 and 224 detectors, which observe at 140 and 240~GHz with resolutions of 18.5 and 12.5~arcsec, respectively. Due to the characteristic spectral distortion of the CMB photons induced by the tSZ effect, {\it NIKA} is an ideal instrument for high resolution tSZ observations. Indeed, the tSZ signal is strongly negative at 140~GHz and positive but close to zero at 240~GHz. The {\it NIKA} prototype has already been successfully tested during four observation campaigns \citep{NIKA_2010,NIKA_2011} at the Institut de Radio Astronomie Millim\'etrique (IRAM) 30-meter telescope at Pico Veleta, Granada, Spain. These observations have demonstrated performances comparable to state-of-the-art bolometer arrays operating at these wavelengths, such as {\it GISMO} \citep{gismo}. The final camera, {\it NIKA2}, will contain 1000 and 4000 detectors at 140 and 240~GHz, respectively, and should be operational in 2015.

We report the first observation of a galaxy cluster via the tSZ effect here using the {\it NIKA} prototype. It has been imaged during the fifth observation campaign of {\it NIKA} in November 2012. The targeted source is the massive intermediate redshift galaxy cluster \mbox{RX~J1347.5-1145} at $z = 0.4516$. It has been selected for both its tSZ intensity and angular size with the latter being comparable to the field of view of the {\it NIKA }prototype. Moreover, \mbox{RX~J1347.5-1145} is known to be a complex merging system that we aim at characterizing further with respect to previous works at scales in the range of 20 to 200~arcsec.
   
This paper is organized as follows. In Sect.~\ref{sec:previous_obs}, we give the status of the previous observations of \mbox{RX~J1347.5-1145}. In Sect.~\ref{sec:observations}, we provide a brief description of the {\it NIKA} camera and give an overview of the observations that is carried out during the November 2012 campaign at the IRAM 30-meter telescope. Sect.~\ref{sec:SZ_analysis} describes the tSZ dedicated data analysis and its validation on simulations is reported in Sect.~\ref{sec:simu_and_idcs}. We present the map of \mbox{RX~J1347.5-1145} in Sect.~\ref{sec:results} and the results on the pressure profile for this cluster of galaxies. These results are then compared to other experiments in Sect.~\ref{sec:comparison}. Throughout this paper, we assume a flat $\Lambda$CDM cosmology according to the lastest {\it Planck} results \citep{PLANCK_param_cosmo} with $H_0 = 67.11$ km.s$^{-1}$.Mpc$^{-1}$, $\Omega_M = 0.3175$, and $\Omega_{\Lambda} = 0.6825$.

%% file: 02_previous_obs.tex
The object RX~J1347.5-1145 is among the clusters that have been intensively observed at several wavelengths and the most widely studied using tSZ at sub-arcmin resolution. It is a massive intermediate redshift galaxy cluster at $z = 0.4516$ undergoing a merging event.

This cluster is the most luminous \mbox{X-ray} cluster of galaxies known to date \citep[{\it e.g.}][]{allen_2002}. It was discovered in the {\it ROSAT} \mbox{X-ray} all-sky survey \citep{voges1999} and has been the object of many studies in \mbox{X-ray} \citep{schindler_1995, schindler_1997, allen_2002, gitti_2004, gitti_2005, gitti_2007, gitti_2007_bis, ota_2008}, optical \citep{cohen_kneib_2002, verdugo_2012}, infrared \citep{zemcov2007}, tSZ \citep{pointecouteau_1999, komatsu1999, pointecouteau_2001, komatsu_2001, kitayama_2004, mason_2010, korngut_2011, zemcov2012, plagge_2012}, and multiwavelength analysis \citep{bradac_2008, miranda_2008, johnson_2012}. From {\it ROSAT} \mbox{X-ray} observations, this cluster was thought to be a dynamically old relaxed cool-core cluster with an extremely strong cooling flow, due to its very spherical morphology and peaked \mbox{X-ray} profile \citep[{\it ROSAT};][]{schindler_1995, schindler_1997}. However, high angular resolution tSZ observations have proved RX~J1347.5-1145 to be an ongoing merger due to the measurement of an extension toward the southeast (SE) with respect to the \mbox{X-ray} center \citep{pointecouteau_1999, komatsu_2001, kitayama_2004}. This illustrates how tSZ and \mbox{X-ray} (and other wavelengths) observations are complementary. More recent \mbox{X-ray} \citep[{\it Chandra};][]{allen_2002} and lensing \citep{miranda_2008} observations are consistent with this interpretation and show a clear detection of the SE extension.

High resolution tSZ maps of RX~J1347.5-1145, such as the 90~GHz 8~arcsec (smoothed to 10~arcsec) resolution map of {\it MUSTANG} \citep{mason_2010}, have confirmed the presence of a strong SE extension. It is interpreted as being due to a hot gas that is heated by the merging of a subcluster crossing the main, originally relaxed, system from the south to the northeast (NE), which is perpendicular to the line-of-sight. The SE extension coincides with a radio mini-halo~\citep{gitti_2007_bis}, which indicates the presence of non-thermal electrons, that underlies a non-thermal contribution to the total pressure. Optical observations have also confirmed this scenario with the detection of a massive elliptical galaxy, which is located 20~arcsec on the east side of the \mbox{X-ray} center, while the central elliptical galaxy of the main cluster remains at the \mbox{X-ray} peak location \citep{cohen_kneib_2002}. 

The temperature profile of RX~J1347.5-1145 varies from $\sim$~6~keV in its core to $\sim$~20~keV at 80~arcsec and decreases to $\sim$~9~keV on the outer part of the cluster (120--300~arcsec form the core). The maximum temperature is located at the SE extension, reaching $k_B T_e \sim~25$~keV \citep{ota_2008}. The Compton $y$ parameter has been measured to be $y_{\mathrm{max}} \simeq~10^{-3}$ \citep{pointecouteau_1999}.

The object RX~J1347.5-1145 hosts a well-known radio source within 3~arcsec of the \mbox{X-ray} center in the central elliptical galaxy. Due to this contamination, the location of the tSZ maximum is still debated. Current single dish observations are consistent with the tSZ emission of the SE extension being stronger than that at the cluster \mbox{X-ray} center. However, taking advantage of the intrinsic point source removal power of interferometric data, \cite{plagge_2012} claim that it is only a secondary maximum. The point source has to be taken into account in the tSZ analysis. According to~\cite{pointecouteau_2001}, the source follows the spectrum $F_{\nu} = \left(77.8 \pm 1.7\right) \nu_{\mathrm{GHz}}^{-0.58 \pm 0.01}$ mJy. For {\it NIKA}, this corresponds to $4.4 \pm 0.3$ and $3.2 \pm 0.2$ mJy at 140 and 240~GHz, respectively.

Finally, in addition to the central radio source, \cite{zemcov2007} have reported the presence of two infrared galaxies. The first one (Z1 hereafter) is located at about 60~arcsec from the \mbox{X-ray} center in the southwest direction with a flux of  15.1 mJy (as measured with a signal to noise of 5.1) at 850 $\mu$m and 125 $\pm$ 34  mJy at 450 $\mu$m. The second source (Z2 hereafter) is located closer to the \mbox{X-ray} center at about 20~arcsec on the northeast side. However, it is only detected at 850 $\mu$m with a flux of 11.4 mJy (measured with a signal to noise of 4.7). The best-fit value at 450 $\mu$m is 10  $\pm$ 32 mJy. The contamination of these sources in the {\it NIKA} bands is estimated and accounted for in our analysis, as discussed in Sect.~\ref{sec:2band_decor}.

%% file: 03_nika_obs.tex
\subsection{Brief overview of the {\it NIKA} camera during the campaign of November 2012}
\label{sec:run_overview}
\begin{table*}
	\begin{center}
	\begin{tabular}{cccc}
	 \hline
         \hline
	& Nov. 21$^{\mathrm{st}}$ & Nov. 22$^{\mathrm{nd}}$ & Nov. 23$^{\mathrm{rd}}$\\
	\hline
	$\tau_{140 \ \mathrm{GHz}}$ & 0.14 & 0.18 & 0.053 \\
	$\tau_{240 \ \mathrm{GHz}}$ & 0.17 & 0.22 & 0.046 \\
	Time range & 8:27 am to 11:43 am & 8:16 am to 12:01 pm & 8:11 am to 10:59 am \\
	Integration time & 2 hrs 29 min & 3 hrs 00 min & 2 hrs 29 min \\
	Unflagged integration time & 50 min & 2 hrs 35 min & 2 hrs 23 min \\
	\hline
	\end{tabular}
	\end{center}
	\caption{Mean zenith opacity, on-source integration time, and period of the day for the three days of the campaign of November 2012. The total integration time is 5 hrs 47 min. The mean opacity ratio is $\tau_{240~\mathrm{GHz}}~/~\tau_{140~\mathrm{GHz}}~\simeq~1.2$}
	\label{tab:table_obs}
	\end{table*}

The {\it NIKA} camera consists of two arrays of Kinetic Inductance Detectors (KIDs) with maximum transmissions at $140$ and $240$~GHz. Ninety percent of the total transmission of the {\it NIKA} bandpasses (see Fig.~\ref{fig:bandpass}) is in the range 127--171~GHz for 140~GHz and 196--273~GHz for 240 GHz bands. The respective angular resolutions (FWHM) are $18.5$ and $12.5$~arcsec with effective fields of view of $1.8 \times 1.8$ and $1.0 \times 1.0$~arcmin. The pitch between pixels is 2.3~mm at 140~GHz and 1.6~mm at 240~GHz. This corresponds to an effective focal plane sampling of 0.77~F$\lambda$ and 0.8~F$\lambda$ at 140 and 240~GHz, respectively. In this particular campaign, the first band ($140$~GHz) was used with $127$ detectors having a mean effective sensitivity of $29$ mJy s$^{1/2}$ per beam ($19$ mJy s$^{1/2}$ per beam for the best $20$\% of all pixels), and the second band ($240$~GHz) had $91$ detectors with a mean effective sensitivity of $55$ mJy s$^{1/2}$ per beam ($37$ mJy s$^{1/2}$ per beam for the best $20$\% of all pixels). This unexpected poor sensitivity and the small number of available detectors for the 240~GHz band is due to the dysfunction of a cold amplifier during this observation campaign. Using only eight detectors of the 240~GHz array, we obtain the expected mean effective sensitivity measured to be $22$ mJy s$^{1/2}$ per beam. Despite the constant improvement in sensitivity over the the last campaigns~\citep{NIKA_2010,NIKA_2011,RFdIdQ}, the sensitivity of the instrument was limited by detector correlated noise coming from electronic and sky noise residuals. For the averaged background during observations, the expected photon noise is 5 mJy s$^{1/2}$ at 140~GHz and 7 mJy s$^{1/2}$ at 240~GHz.

Unlike traditional bolometric instruments, {\it NIKA} uses KIDs. The KIDs are superconducting resonators whose resonance frequency ($\sim$ 1--2.5~GHz) changes linearly with the absorbed optical power \citep[see for example][]{swenson_2010}. Each resonator can be modeled by a complex transfer function in frequency with a real part $I$ (in-phase) and imaginary part $Q$ (quadrature) \citep{grabovskij_2008}. By measuring $I$ and $Q$ at a constant frequency (defined for each detector by the electronics) as a function of time, we can reconstruct the shift of the resonance frequency, as described in \citet{RFdIdQ}. This method allows us to obtain accurate photometry to be better than 10\%.

The KIDs used here are Hilbert dual-polarization designed LEKID pixels \citep[Lumped Element KID;][]{doyle_2008, roesch_2012}, which are realized on 180 $\mu$m and 275 $\mu$m thickness silicon substrate at 240 and 140 GHz, respectively. The detector resistivity is larger than 5000 $\Omega$ cm for both wavelengths. The detectors are cooled down to about 100~mK with a $^4\mathrm{He}$ -- $^3\mathrm{He}$ dilution cryostat. 

More details on the {\it NIKA} prototype setup can be found in~\cite{main_run5}.

\subsection{Observing strategy of the targeted galaxy clusters}
\label{sec:obs_condition}
Galaxy clusters are weak extended sources when seen through the tSZ effect, making their observations challenging.
For this study, we have selected \mbox{RX~J1347.5-1145}, which is an intermediate redshift cluster at $z = 0.4516$.
 The object \mbox{RX~J1347.5-1145} is among the most luminous tSZ sources in the sky, and
 it is also compact enough to have an angular size comparable to the field of view of the NIKA camera. 

\begin{figure}
	\centering
	\includegraphics[width=\columnwidth]{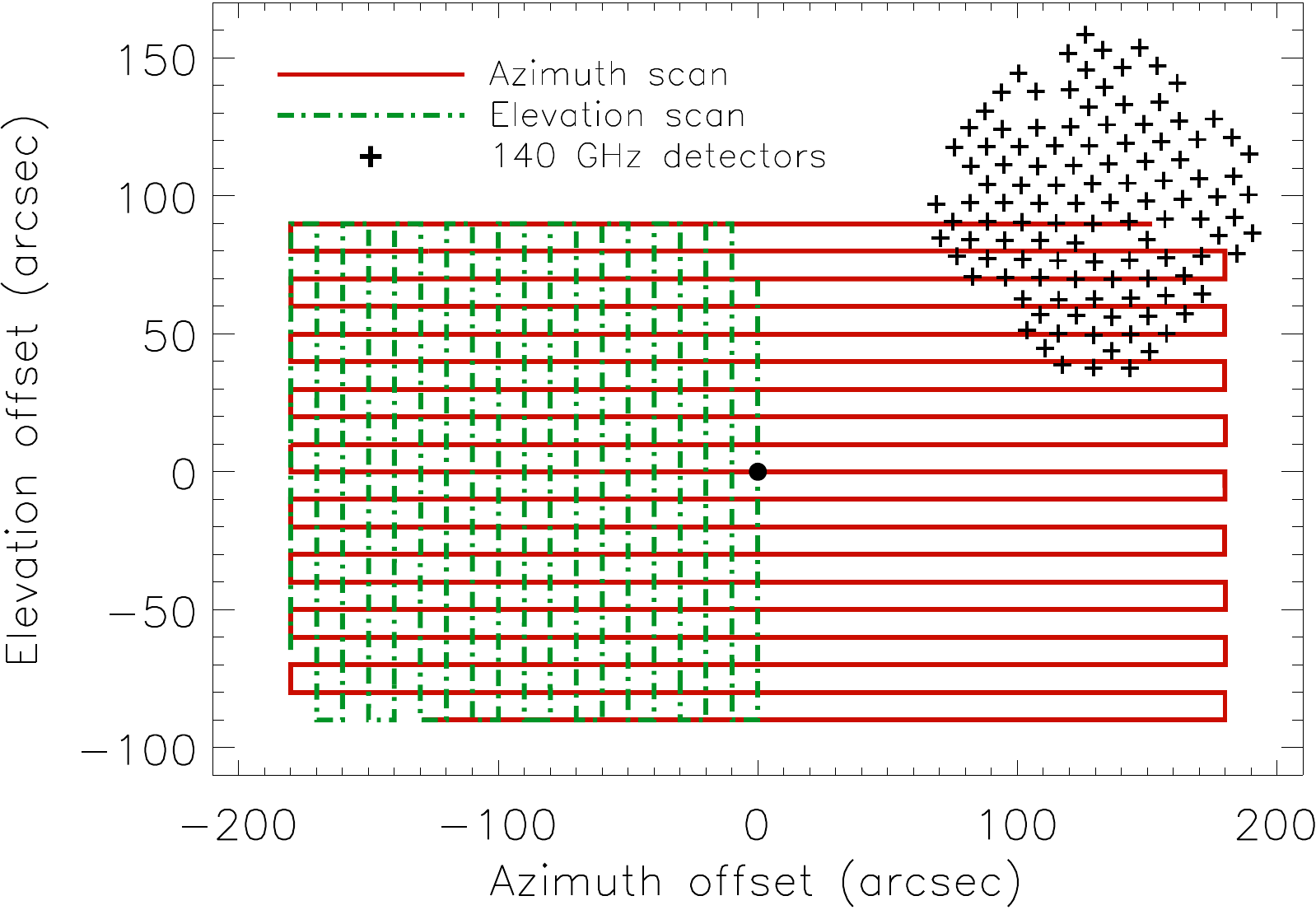}
	\caption{Elevation (dashed green) and azimuth (solid red) offset scans. The center is represented by a black dot and has coordinates (R.A.,~Dec)~=~(13h~47m~32s,~-11$^{\mathrm{o}}$~45'~42"). The 140~GHz array is also represented by black crosses, which correspond to the position of each KID in the focal plane (gaps in the array correspond to invalid detectors).}
        \label{fig:scans}
	\end{figure}

As shown in Fig.~\ref{fig:scans}, the cluster signal is scan-modulated but there is no wobbling involved. Raster scans are made of constant elevation subscans or constant azimuth subscans. For the latter, only the low azimuth part of the field was covered due to an error in the control software. Both of them are $6$ min $20$s scans that are made of $19$ subscans separated by $10$~arcsec steps. Scans along the azimuth direction are centered at (R.A.,~Dec)~=~(13h~47m~32s,~-11$^{\mathrm{o}}$~45'~42"), which sample a rectangular region of $360 \times 180$~arcsec (azimuth $\times$ elevation), while scans along the elevation sample a region of $180 \times 180$~arcsec and are centered on a point $90$~arcsec away from (13h~47m~32s,~-11$^{\mathrm{o}}$~45'~42"), which rotates with the parallactic angle. The scan velocity is about 15~arcsec s$^{-1}$. The detailed integration times are given in Table~\ref{tab:table_obs} with the corresponding atmospheric opacities.
	
\subsection{Pointing, calibration, bandpasses, and beam}
\label{sec:calib_pointing}
Uranus observations were used to reconstruct beam maps (projection of the array on the sky and measure of individual detector beams) for both wavelenghts. Nearby quasars were used for determining pointing corrections. The pointing root mean square error is estimated to be $\sim$3~arcsec \citep{main_run5}. This is small compared to the beam and has a negligible impact in the case of extended sources such as \mbox{RX~J1347.5-1145}.

We also used Uranus for absolute point source flux calibration. The flux of the planet was inferred from a frequency dependent model of the planet brightness temperature taken from \cite{moreno2010}. The Uranus brightness temperatures are typically 113~K at 140~GHz and 94~K at 240~GHz. This model is integrated over the {\it NIKA} bandpasses for each channel, and it is assumed to be accurate at the 5\% level. The final absolute calibration factor is obtained by fitting the amplitude of a Gaussian function of fixed angular size on the reconstructed maps of Uranus (representing the main beam). We neglect the angular diameter of Uranus, 3.54~arcsec at the time of the observations, when it is compared to the size of the main beam, since the convolution of the corresponding disk with a Gaussian of 12.5 and 18.5~arcsec full width at half maximum (FWHM) broadens our beam by only 0.17~arcsec at 240~GHz and 0.12~arcsec at 140~GHz.

Scales larger than 180~arcsec, which correspond to the scan size, were not measured with {\it NIKA}. By integrating the Uranus flux up to 100~arcsec, we observe that the total solid angle covered by the beam, which includes the power in the side lobes, is larger than the Gaussian best-fit of the main beam by a factor of 1.32. Scales larger than 100~arcsec are noise dominated on the Uranus map. Thus, using recent measurements of the IRAM 30-meter beam pattern with EMIR \citep{K13}, we extrapolate the angular profile of the beam from 100~arcsec to 180~arcsec, and find an overall factor equal to 1.45 \citep[see][for a more detailed description]{main_run5}. From the dispersion over different observations of Uranus, we estimate the uncertainties on the solid angle of the main beam to be about 4 \%.  We obtain 10 \% uncertainties for the full beam by also considering uncertainties on the side lobes. 

The sky maps (also for Uranus maps prior to calibration) are corrected for atmospheric absorption using elevation scans, or skydips \citep[see][for further details]{main_run5}. In our case, the resonance frequencies of the detectors are measured versus the optical load, which depends on the zenith opacity and the elevation. This gives the zenith opacity as a function of the resonance frequency of the detectors, which is measured for each scan. The opacity can then be corrected to good accuracy by accounting for the air mass at the elevation of the source. Furthermore, different atmospheric conditions lead to changes in the beam pattern of the instrument that also affect the absolute calibration accuracy \citep{main_run5}. From the dispersion of the recovered flux of Uranus, which was observed several times with different opacities during the campaign, we estimate an overall accuracy of 15\% \citep{main_run5} for the calibration procedure.

To summarize, the list of the main systematic uncertainties in the 140~GHz band are listed in Table~\ref{tab:table_err}. The total calibration uncertainty on the final data at the map level is estimated to be 16\%.
\begin{table}
\begin{center}
\begin{tabular}{cc}
\hline
\hline
Systematic uncertainty & Error percentage \\
\hline
Brightness temperature model & 5\% \\
Point source calibration & 15\% \\
Secondary beams fraction & 45\% $\pm$ 10 \% \\
Bandpasses & 2\% \\
\hline
\end{tabular}
\end{center}
\caption{Main contributions to the absolute error of the {\it NIKA} data for the 140 GHz band.}
\label{tab:table_err}
\end{table}

%% file: 04_sz_ana.tex
\subsection{Thermal Sunyaev-Zel'dovich data}
\label{sec:tsz_data}
    	\begin{figure}
	\centering
	\includegraphics[width=\columnwidth]{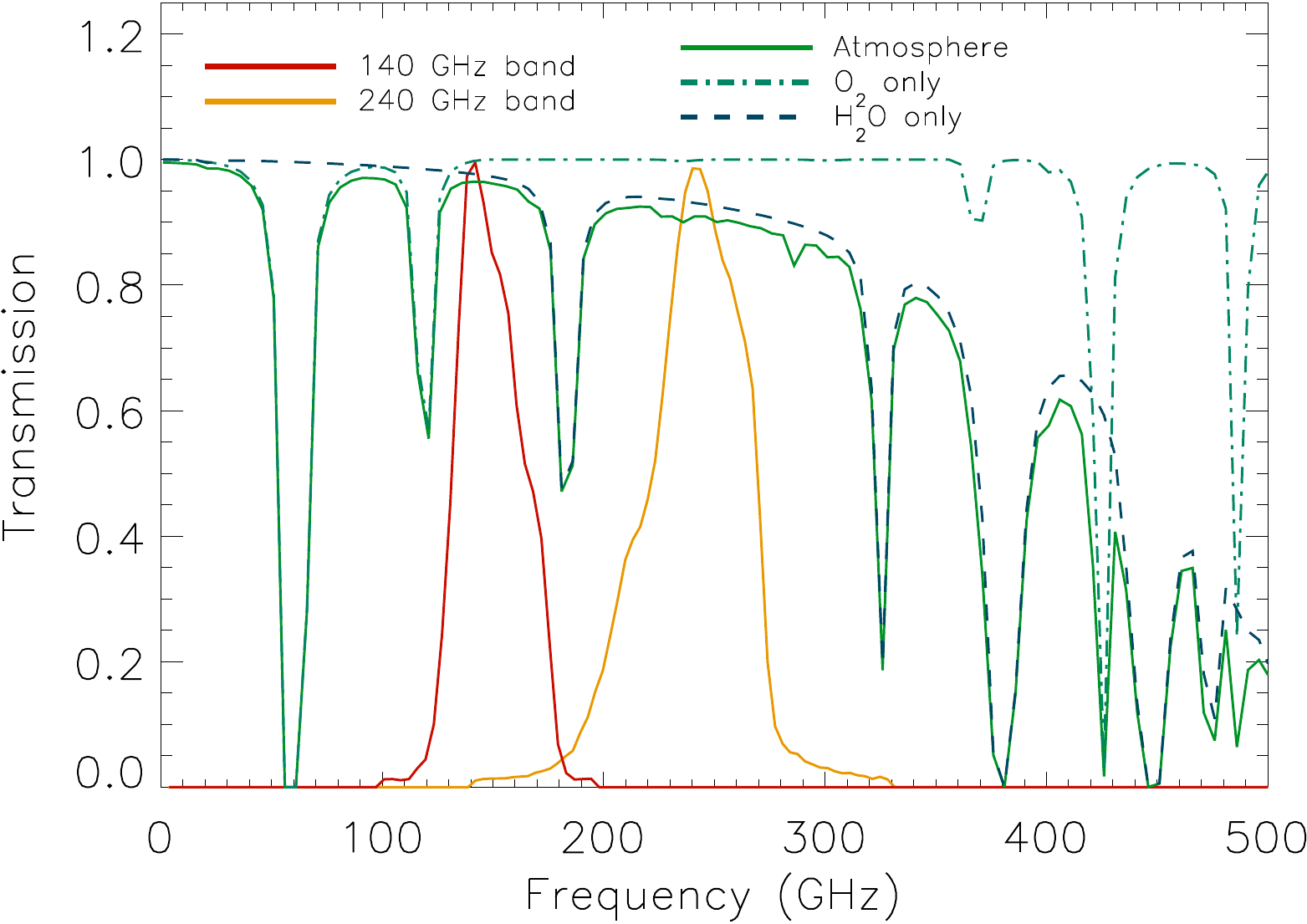}
	\caption{Normalized 140~GHz (solid red line) and 240~GHz (solid orange line) instrumental bandpasses. The total atmospheric transmission is also given as a solid green line for 1~mm of precipitable water vapor, according to the Pardo model \citep{Pardo}. The oxygen (dash-dotted light blue) and the water vapor (dashed dark blue) contributions are represented.}
        \label{fig:bandpass}
	\end{figure}
	
In the non-relativistic limit, the tSZ effect results in a distortion of the CMB black-body spectrum, whose intensity frequency dependence is given by \citep{Birkinshaw}
\begin{equation}
	g(x) = - \frac{x^4 e^x}{\left(e^x-1\right)^2} \left(4 - x  \ \mathrm{coth}\left(\frac{x}{2}\right) \right),
	\label{eq:sz_f_x}
\end{equation}
where $x = \frac{h \nu}{k_{\mathrm{B}} T_{\mathrm{CMB}}}$ is the dimensionless frequency; $h$ is the Planck constant, $k_{\mathrm{B}}$ the Boltzmann constant, $\nu$ the observation frequency and $T_{\mathrm{CMB}}$ the temperature of the CMB. The induced change in intensity relative to primary CMB intensity $I_0$ reads
\begin{equation}
	\frac{\delta I_{\mathrm{tSZ}}}{I_0} = y \ g(x),
\label{eq:Taniso}
\end{equation}
 where $y$ is the Compton parameter. The latter measures the integrated electronic pressure $P_{\mathrm{e}}$ along the line-of-sight
   \begin{equation}
	y = \frac{\sigma_{\mathrm{T}}}{m_{\mathrm{e}} c^2} \int P_{\mathrm{e}} dl.
	\label{eq:y_compton}
   \end{equation}
The parameter $\sigma_{\mathrm{T}}$ is the Thomson cross section, $m_{\mathrm{e}}$ is the electron mass, and $c$ the speed of light. The tSZ spectral distortion is null at 217~GHz, negative below this frequency, and positive above.

The unit conversion coefficients between Jy/beam and Compton parameter $y$ are $-11.8 \pm 1.2$ and $+ 2.2 \pm 0.6$ at 140 and 240~GHz, respectively, for the {\it NIKA} prototype. These coefficients are computed by taking the overall transmission of the instrument and the measured total beam with their respective errors into account. The Compton parameter $y$ is first converted to Jy/sr using Eq.~\ref{eq:Taniso} and then converted to Jy/beam using the angular coverage of the beam. We assume a pure non-relativistic tSZ spectrum.

As the expected tSZ signal is small (up to $\sim$ 10 mJy/beam), the {\it NIKA} raw data are dominated by instrumental noise and atmospheric emission. We model the signal measured by a KID $k$, which operates at the observing frequency band $\nu_b$ (140~GHz or 240~GHz) as
   \begin{equation}
	d_k(\nu_b, t) = S_k(\nu_b, t) + N_k(t) + E(\nu_b, t) + A(\nu_b, t).
	\label{eq:signal_and_noise}
   \end{equation}
The astrophysical signal (essentially tSZ) $S_k(\nu_b, t)$ is time-dependent through the scanning strategy. Furthermore, it varies with the frequency band (Eq.~\ref{eq:Taniso}) and with the detector $k$ because of its location in the focal plane. The variable $N_k(t)$ is the uncorrelated detector noise limiting the sensitivities given in Sect.~\ref{sec:run_overview}. The correlated electronic noise, $E(\nu_b, t)$, is well characterized by an identical common-mode for the detectors of the same band \citep{NIKEL}. As we use independent readout electronics for the two bands, the electronic noise is uncorrelated between bands. Finally, by splitting the frequency and time dependance, the atmospheric contribution can be modeled as
   \begin{equation}
   	\begin{array}{lcr}
	A(\nu_b, t) = a_{\mathrm{H}_2\mathrm{O}} ^{\mathrm{el}} (\nu_b) \ A_{\mathrm{H}_2\mathrm{O}}^{\mathrm{el}} (t) \ + \ a_{\mathrm{O}_2} ^{\mathrm{el}} (\nu_b) \ A_{\mathrm{O}_2}^{\mathrm{el}} (t) \\
\\
\hspace*{1.15cm}	+ \  a_{\mathrm{H}_2\mathrm{O}} ^{\mathrm{fluc}} (\nu_b) \  A_{\mathrm{H}_2\mathrm{O}}^{\mathrm{fluc}} (t).\\
	\end{array}
	\label{eq:atmosphere}
   \end{equation}
   The first and the second terms give the emission change of water vapor and oxygen due to the variation of the airmass with the elevation. The third term, $a_{\mathrm{H}_2\mathrm{O}} ^{\mathrm{fluc}} (\nu_b) \ A_{\mathrm{H}_2\mathrm{O}}^{\mathrm{fluc}} (t)$, gives the emission change due to inhomogeneities in the water vapor distribution. We note that $a_{\mathrm{O}_2}^{\mathrm{fluc}}$ is implicitly set to zero because of the assumption that the oxygen is locally very homogeneous in the atmosphere. It is also important to notice that the two bands are not sensitive to the same atmospheric components. The 140~GHz band is sensitive to the $\mathrm{O}_2$ 118~GHz line, while the 240~GHz band is almost only sensitive to water vapor \citep{Pardo}, such that $\frac{a_{\mathrm{O}_2} ^{\mathrm{el}} (140 \ \mathrm{GHz})}{a_{\mathrm{H}_2\mathrm{O}} ^{\mathrm{el}} (140 \ \mathrm{GHz})} \gg \frac{a_{\mathrm{O}_2} ^{\mathrm{el}} (240 \ \mathrm{GHz})}{a_{\mathrm{H}_2\mathrm{O}} ^{\mathrm{el}} (240 \ \mathrm{GHz})}$. This can be observed in Fig.~\ref{fig:bandpass}, where we show the bandpasses of the {\it NIKA} prototype in red (140~GHz) and orange (240~GHz). The atmospheric transmission is given for the oxygen (light blue dash-dotted line) and water vapor ( dark blue dashed line) contributions. The overall atmospheric transmission is given as a green solid line, according to the Pardo model \citep{Pardo}. Trace constituents are neglected here ({\it e.g.}, ozone).

\subsection{Time ordered data analysis}
 \label{sec:TOI_ana}
 	\begin{figure*}
	\centering
	\includegraphics[height=6cm]{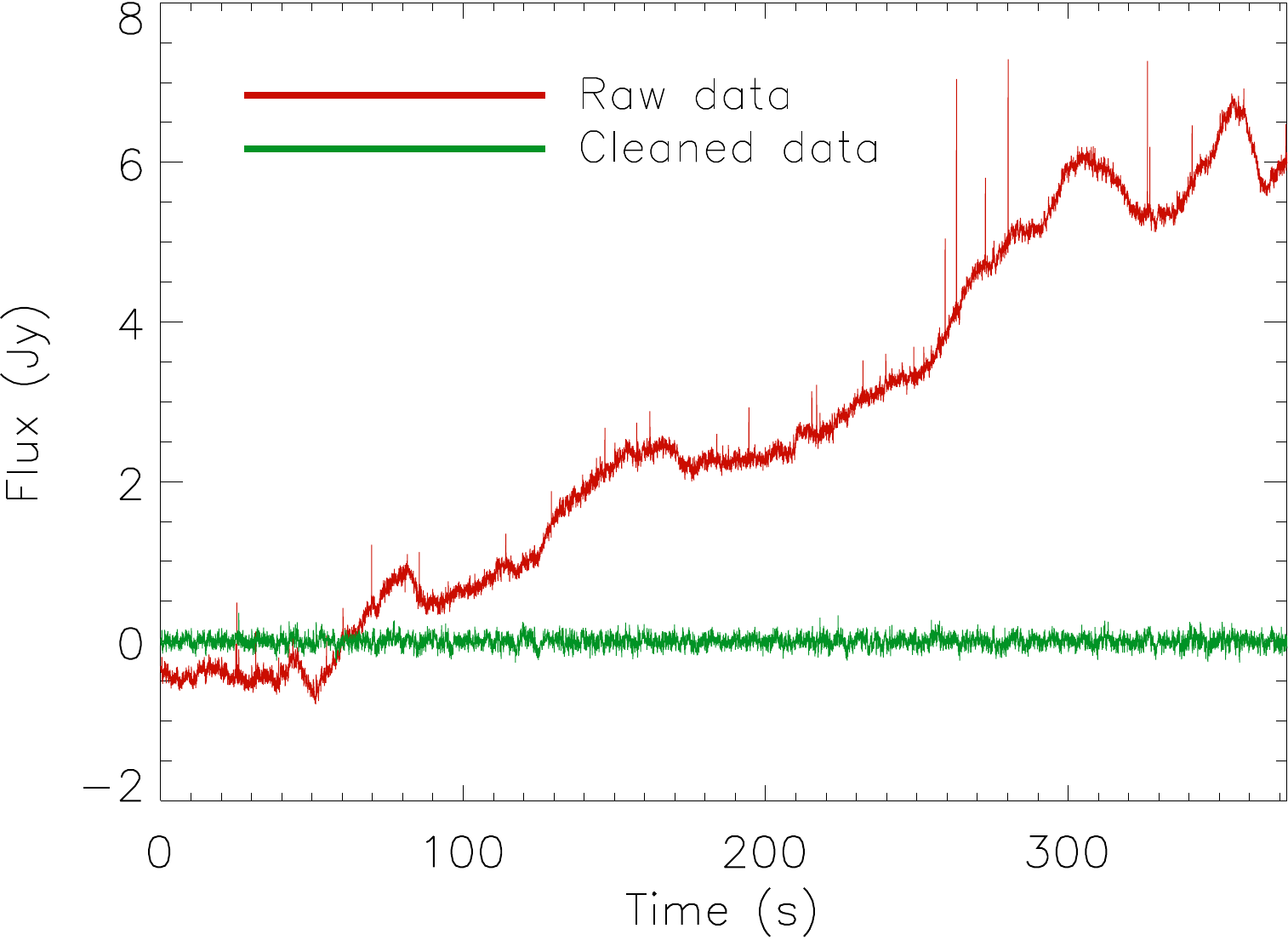}
	\hspace*{0.5cm}
	\includegraphics[height=6cm]{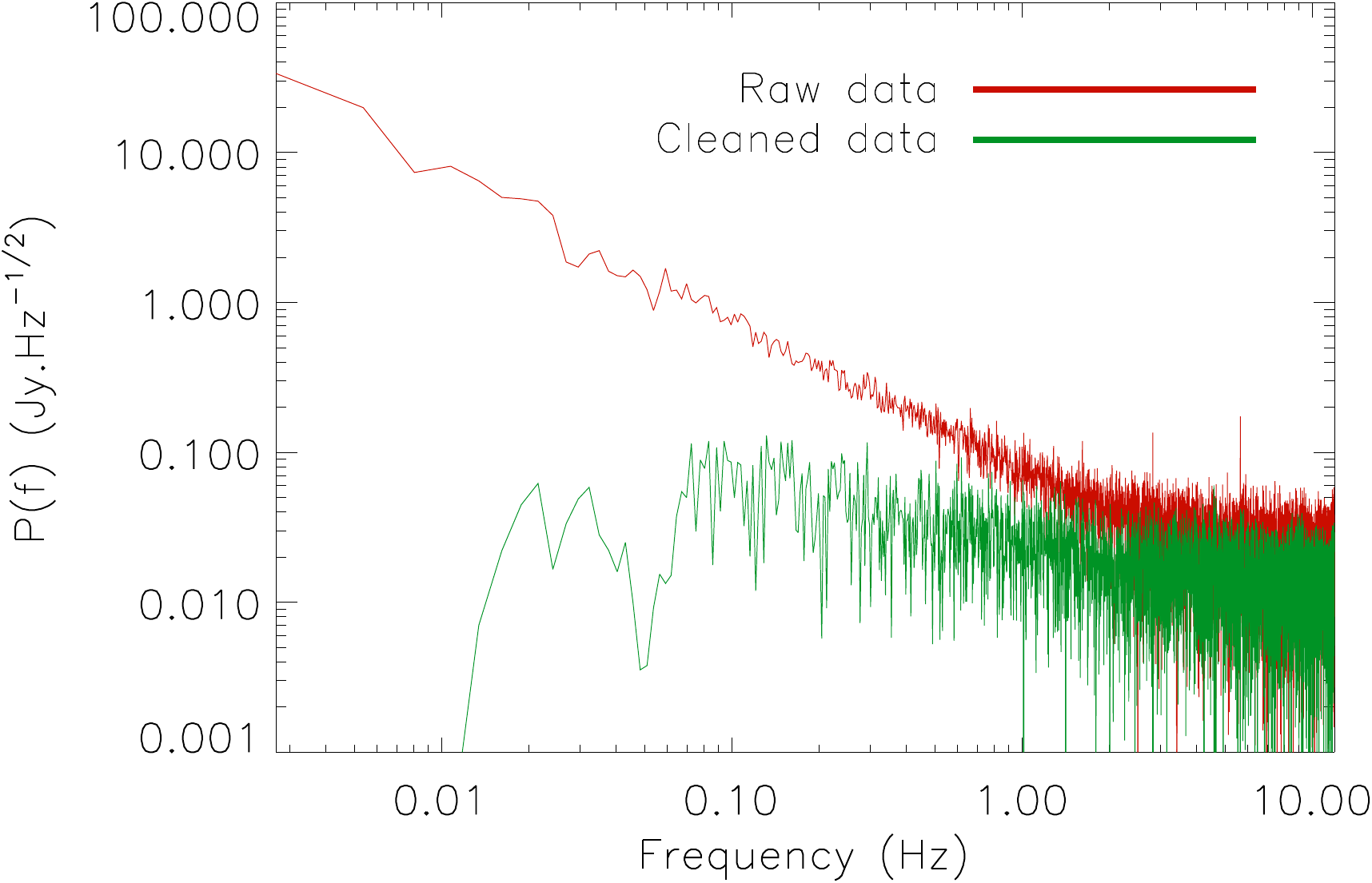}
	\caption{TOD (left) and their power spectra (right) for a given detector. The data corresponds to the calibrated TOD before (red) and after (green) the electronic and atmospheric noise decorrelation. The TOD are dominated by the atmospheric noise at low frequencies, which is responsible for the slow variations in the red TOD and the obvious rise of noise below $\sim 1$~Hz on the red power spectrum. Cosmic rays hitting the instrument can be seen as spikes in the TOD but have been removed before computing the power spectra. Pulse tube frequency lines appear in the power spectrum ({\it e.g.} the $\sim 6$~Hz line in the raw power spectrum) and are notch filtered. The electronic noise dominates at frequencies between $\sim 1$ and $\sim 5$~Hz in the power spectrum before decorrelation.} \label{fig:TOI_PS_real}
	\end{figure*}
The main steps for processing the time ordered data (TOD) are listed below.
\begin{itemize}
\item Loading raw data, including the telescope parameters, the reconstruction of the projection of the array on the sky, and the atmospheric opacity.
\item Calibrating the TOD, including opacity correction.
\item Flagging invalid detectors.
\item Flagging cosmic ray impacts on the detectors.
\item Decorrelating atmospheric and electronic noise.
\item Filtering low frequencies and removing lines produced by the pulse tube of the cryostat with a notch filter.
\item Making map using inverse variance weighting.
\end{itemize}
In the following, we give details on specific points of the analysis.

\subsubsection{Raw data}
The raw TOD correspond to the real ($I_k(t)$, in-phase) and imaginary ($Q_k(t)$, quadrature) parts of the transfer function of the system (array and transmission line), which are sampled on predefined frequency tones $k$ at an acquisition rate of 23.842~Hz. We also compute the average modulation of these quantities with respect to the injected frequency (typically a few~kHz), which are noted $\delta I_k(t)$ and $\delta Q_k(t)$. These four quantities are used to reconstruct the shift of the resonance frequency ${\delta f_0}_k (t)$, which probes the optical power absorbed by a detector \citep[see][for more details]{RFdIdQ}. To monitor the electronic noise and possible variations of the transfer function of the transmission line, the latter is also sampled with tones that are placed off-resonance (with no correspondence to any detector), which are insensitive to optical power.

In the case of the {\it NIKA} prototype, some detectors are subject to cross talk and are not used for this analysis. Bad detectors are also flagged on the basis of the statistical properties of their noise. In particular, we use skewness and kurtosis tests, in addition to testing the stationarity of the noise. Some TODs, which are affected by baseline jumps due to the coupling with ambient magnetic fields, are also excluded. These rejected detectors are not used in the following. For the observations of \mbox{RX~J1347.5-1145}, the number of detectors used in the analysis is 81 at 140~GHz and 45 at 240~GHz.

\subsubsection{Calibration}
The shift of the resonance frequency is computed for each detector. The absolute calibration from resonance frequency to flux density is applied to these TODs. The beam is measured with Uranus observations in atmospheric conditions that are similar to those for the \mbox{RX~J1347.5-1145} observations. The Uranus data are fitted with a Gaussian function of FWHM that is equal to 12.5 and 18.5~arcsec at 240 and 140~GHz, respectively. An opacity correction is performed by multiplying the data by $\mathrm{exp}\left( \tau_{\nu_b} / \mathrm{sin}(el) \right)$, where $el$ is the elevation of the source. The calculation of the opacity is based on skydip measurements, as briefly described in Sect.~\ref{sec:calib_pointing} \cite[for more details, see][]{main_run5}.

\subsubsection{Glitch removal}
Cosmic rays hitting the instrument induce glitches in the data. The time response of KIDs is negligible compared to the sampling frequency, such that a cosmic ray impact appears as a peak on a single data sample in the TOD. We detect about four glitches per minute. They are removed from the ${\delta f_0}_k (t)$ TODs by flagging peaks that are above five times the standard deviation of the considered TOD. The TODs are flagged and interpolated at the glitch locations in order not to affect the decorrelation. These flagged data are not projected onto maps. 

\subsubsection{Dual-band decorrelation}
\label{sec:2band_decor}
As discussed above the atmospheric contribution to the NIKA data, $A(\nu_b, t)$, is essentially due to water vapor, to first order. Therefore, it is expected to be the same for the two frequency bands up to an amplitude factor $A(240~\mathrm{GHz},~t)/A(140~\mathrm{GHz},~t)~\simeq~5$. As a consequence, we first use the 240~GHz data to build an atmospheric template and remove it from the 140~GHz data by linear fitting. The fit is performed  for each subscan and for each 140~GHz detector independently. As the tSZ signal at 240~GHz is smaller by a factor of 5.5 (see  unit conversion factors between Compton parameter and Jansky per beam  in Sect.~\ref{sec:tsz_data}) with respect to the one at the 140~GHz band, the positive bias introduced in the 140~GHz data by the tSZ signal present at 240~GHz is negligible (about 27 times smaller). 

As shown in the left panel of Figure~\ref{fig:TOI_PS_real}, where we present the raw TOI signal for a typical KID at 140~GHz in red, the atmospheric contribution is dominated by a low frequency component. These drifts correspond to the $1/f$ noise-like spectrum with a knee frequency at about 1~Hz, which is presented in red on the right panel of the figure. We thus apply a low-pass filter to the atmospheric template that is deduced from the 240~GHz channel. This filter removes most of the detector correlated electronic noise in the 240~GHz band, $E(240 \ \mathrm{GHz}, t)$, which does not affect the 140~GHz data.  Furthermore, it allows us to reduce the intrinsic high noise level of the 240~GHz band, which was specific to the November 2012 {\it NIKA} data due to the cold amplifier disfunction and would otherwise pollute the 140~GHz cleaned data. The low-pass filter does not affect frequencies smaller than 1.5~Hz and sets frequencies larger than 2~Hz to zero. Frequencies between 1.5 and 2~Hz are progressively attenuated using a cosine squared cutoff. 

We also build a template from a high-pass filtered common-mode obtained from the TODs of the valid 140GHz detectors. This high-pass filter is the complement to the previous low-pass filter such that their sum is equal to one for all frequencies. We linearly fit this template to each subscan of each 140GHz detector TOD and remove it. As a consequence the correlated electronic noise, $E(140 \ \mathrm{GHz}, t)$ is removed at frequencies larger than the cutoff. We note that this does not significantly affect the tSZ signal because it is not correlated at these high frequencies between detectors as they observe different positions on the sky. Typically, 2~Hz corresponds to about 8~arcsec for the chosen scan speed (about 15~arcsec per second).

Finally, we fit and remove a template that follows the elevation of the telescope from the TODs. Indeed, as the oxygen component of the atmosphere is ignored in the dual-band decorrelation, it appears as a residual proportional to the elevation of the telescope.  \\

The main drawback of the dual-band decorrelation technique is the possible contamination of the 140~GHz tSZ reconstructed signal by other astrophysical components present at 240~GHz. First, we consider the kinematic Sunyaev-Zel'dovich (kSZ) effect, which is due to the overall motion of a cluster (or its components) with respect to the CMB reference frame and follows a pure black-body spectrum at the CMB temperature. The kSZ signal is also reduced by a factor of $\sim 5$, such that its flux at 240~GHz should be larger than half the tSZ flux at 140~GHz to produce a non-negligible bias. This is not the case even for the most extreme clusters, such as \mbox{MACS~J0717.5+3745}~\citep{mroczkowski_2012, sayers2013}. Therefore, any kSZ signal present at 240~GHz is neglected in the following analysis.

We have also searched for dusty galaxies within the cluster or gravitationally lensed submilimeter high-redshift background objects that might also contaminate the signal at 140 and 240~GHz. As mentioned in Sect.~\ref{sec:previous_obs}, two of such sources are present in \mbox{RX~J1347.5-1145}. Despite the high level of noise in the 240~GHz {\it NIKA} data for this campaign, the first source, Z1, (R.A,~Dec)~=~(13h~47m~27.6s,~-11$^{\mathrm{o}}$~45'~54") is observed in this band with a flux of 12.7 mJy $\pm$ 6.2. The second source, Z2, (R.A,~Dec)~=~(13h~47m~31.3s,~-11$^{\mathrm{o}}$~44'~57") is not detected in the 240~GHz {\it NIKA} band, but we obtain an upper limit of 4.4 mJy at 1 $\sigma$. Using this information and the measured fluxes at 850 and 450~$\mu$m (see Sect.~\ref{sec:previous_obs}), we estimate the expected flux of the sources at 140~GHz. Assuming dust temperatures in the range of 15 -- 20 K and dust spectral indexes $\beta_{\mathrm{d}}$ in the range of 1.5 -- 2, we obtain 0.85 mJy  at 140~GHz for the Z1 source. For the second source, Z2, we are not able to fit a typical dust spectrum and only compute an upper limit of the flux at 140~GHz by assuming a Rayleigh-Jeans spectrum at low frequency and using the 240~GHz estimated flux. We obtain an upper limit at 140~GHz of 0.65 mJy. In the context of dual-band decorrelation, the 240~GHz fluxes are scaled down by $\simeq 5$, and they are diluted by another factor of $\simeq 5$ due the averaging over all the 240~GHz detectors, which observe the source at different time samples. As uncertainties on the estimated flux for both sources and at both wavelengths are large, we choose not to subtract them directly in the TOD but to account for them in the final analysis, as discussed in Sect.~\ref{sec:ps_sub_effect}.

\subsubsection{Fourier filtering}
Frequency lines ({\it e.g.} at $\sim$ 6~Hz) are induced by the pulse tube of the cryostat and observed in the TOD power spectra (see Fig.~\ref{fig:TOI_PS_real}, right panel). They are flagged and set to zero. In addition, we apply a high-pass filter to remove correlated noise at frequencies lower than the subscan because no tSZ signal is present there. We also remove low frequency (below 0.05~Hz) sine and cosine from the data to further subtract correlated noise contamination.

\subsection{Mapmaking}
\label{sec:mapmaking}
Finally, we construct surface brightness maps by projecting and averaging the signal from all KIDs on a pixelized map at 140~GHz. The projected data are weighted according to the level of noise of each detector using inverse variance weighting. To remove possible offsets in the TOD, we subtract the mean value of each TODs, and we take the zero level as the mean of the external part of the map, where no signal is detected.
	
\subsection{Point source subtraction}
\label{sec:point_source}
The object RX~J1347.51145 hosts a radio point source located within 3~arcsec of the X-ray center that has a flux of $4.4 \pm 0.3$~mJy and $3.2 \pm 0.2$~mJy  at 140 and 240~GHz, respectively~\citep{pointecouteau_2001}. The point source is subtracted in the TODs at both frequencies before the processing, so that it does not bias the analysis. We discuss in Sect.~\ref{sec:ps_sub_effect} how uncertainties on the point source subtraction affect the final results.

%% file: 05_simu.tex
We present two independent validations of the analysis pipeline. The first of them is based on a detailed simulation of the observation of a tSZ cluster with known physical parameters and typical atmospheric and electronic noise. The second one is based on the observation of a faint cluster that allow us to show that the tSZ detection is not an artifact of the data analysis and/or acquisition. 

 \subsection{Simulation}
 \label{sec:simu}	
To test the pipeline as described in Sect.~\ref{sec:SZ_analysis}, we simulate the {\it NIKA} observations of a cluster and construct the TODs by taking all terms of Eq.~\ref{eq:signal_and_noise} into account, which includes the atmospheric contamination, the electronic noise, and the tSZ signal. The parameters used in the simulation are given in Table~\ref{tab:table_param_simu} and are representative of the weather conditions for the observations described in this paper.
\begin{table}
\begin{center}
\begin{tabular}{cc}
\hline
\hline
Parameter & Value \\
\hline
$v_{\mathrm{H}_2\mathrm{O}}$ & 1 \ m/s\\
$h_{\mathrm{H}_2\mathrm{O}}$ & 3000 \ m \\
$\alpha_{\mathrm{H}_2\mathrm{O}}$ & -1.35 \\
$\tau_{140 \ \mathrm{GHz}}$ & 0.1 \\
$\tau_{240 \ \mathrm{GHz}}$ & 0.12 \\
$\left({F_{\mathrm{H}_2\mathrm{O}}}\right)_{140 \ \mathrm{GHz}}$ & 28 \ Jy/beam \\
$\left({F_{\mathrm{H}_2\mathrm{O}}}\right)_{240 \ \mathrm{GHz}}$ & 110 \ Jy/beam \\
$F_{\mathrm{el}} (140 \ \mathrm{GHz})$ & 14 \ Jy/beam/K \\
$F_{\mathrm{el}} (240 \ \mathrm{GHz})$ & 35 \ Jy/beam/K \\
$T_{\mathrm{atmo}}$ & 233 \ K \\
$E_0(1 \ \mathrm{Hz,} \ 140 \ \mathrm{GHz})$ & 38 \ mJy/beam \\
$E_0(1 \ \mathrm{Hz,} \ 240 \ \mathrm{GHz})$ & 76 \ mJy/beam \\
$\beta$ & -0.25 \\ 
$N_0 (140 \ \mathrm{GHz})$ & 29 \ $\mathrm{mJy.s}^{1/2}$ \\
$N_0 (240 \ \mathrm{GHz})$ & 57 \ $\mathrm{mJy.s}^{1/2}$ \\ 
$R_{\mathrm{g}}$ & 0.065 s$^{-1}$ \\
\hline
\end{tabular}
\end{center}
\caption{Values of the parameters used in the simulation; see text for details.}
\label{tab:table_param_simu}
\end{table}

 \subsubsection{Atmospheric contribution}
 \label{sec:atmo_simu}
 The atmospheric contribution $A(\nu_b, t)$ is simulated as described in Sect.~\ref{sec:SZ_analysis}.

The water vapor fluctuations ({\it i.e.}, $a_{\mathrm{H}_2\mathrm{O}} ^{\mathrm{fluc}} \times A_{\mathrm{H}_2\mathrm{O}}^{\mathrm{fluc}} (t)$) are obtained by simulating a map of water vapor anisotropy that passes in front of the telescope aperture with a speed $v_{\mathrm{H}_2\mathrm{O}}$ at an altitude $h_{\mathrm{H}_2\mathrm{O}}$ above the telescope. The power spectrum of this map is a power law with slope $\alpha_{\mathrm{H}_2\mathrm{O}}$. The amplitudes of the atmospheric fluctuations are then normalized to have a standard deviation over the time of the scan equal to $\sigma = F_{\mathrm{H}_2\mathrm{O}} \left(1-\mathrm{exp} \left(-\frac{\tau}{\mathrm{sin} (el)}\right) \right)$, where $\tau$ is the zenith opacity, $el$ the elevation, and $F_{\mathrm{H}_2\mathrm{O}}$ is the reference flux. 

The contribution of the elevation terms, both from $\mathrm{H}_2\mathrm{O}$ and $\mathrm{O}_2$, is simulated as 
   \begin{equation}
	  d_{\mathrm{el}}(t)~=~F_{\mathrm{el}} T_{\mathrm{atmo}} \left(\mathrm{exp}\left(-\frac{\tau}{\mathrm{sin}(<el>)}\right) - \mathrm{exp}\left(-\frac{\tau}{\mathrm{sin}(el)}\right)\right).
	\label{eq:el_term}
   \end{equation}
The parameter $T_{\mathrm{atmo}}$ is the temperature of the atmosphere, and $F_{\mathrm{el}}$ is a reference flux that is measured at both frequencies using skydips.

 \subsubsection{Electronic noise}
 \label{sec:elec_simu}
The electronic noise $E(\nu_b, t)$ is simulated as a common mode with a power spectrum slope $\beta$ and an amplitude $E_0$. This common-mode is identical for all detectors in a given frequency band but differs for the two bands since the electronics is not the same. The spectrum slope is the same for the two bands, but the amplitude is higher at 240~GHz than at 140~GHz (see Table~\ref{tab:table_param_simu}).
 
 \subsubsection{Uncorrelated noise}
 \label{sec:white_simu}
We also simulate a total uncorrelated noise term, $N_k(t)$, independently for each detector. This term accounts for various white noise contributions, including  photon noise, spontaneous Cooper pair breaking due to thermal noise fluctuations and electronic white noise. For the purpose of the simulations, we keep this noise contribution independent of the observing conditions. However, for the real observations, we find that the white noise level is coupled to the atmospheric conditions. This is mainly due to the broadening of the resonances for large optical loads, and it is not accounted for in the simulations. Similarly, we do not account for photon noise variations induced by changes in the optical load. For the sake of simplicity, the total root mean square of the uncorrelated noise is assumed to be identical for all detectors of the same array.
	
\subsubsection{Glitches}
\label{sec:glitch_simu}
Glitches are simulated with a rate $R_{\mathrm{g}}$. They only affect individual samples in the TODs ({\it i.e.}, the KID response is much faster than the sampling frequency) and simultaneously affect all KIDs of a given array ({\it i.e.}, glitches are assumed to generate phonons that hit all the KIDs of the array). The amplitudes of the glitches are generated using a Gaussian spectrum with a standard deviation of 1.3~Jy/beam, as observed in the measured TODs.
		
  \subsubsection{Pulse tube lines}
 \label{sec:ptl_simu}
To simulate the frequency lines generated by the pulse tube, we add cosine functions to the timeline that correspond to the typical frequencies and amplitudes seen in the data (see the power spectrum of the raw data in Fig.~\ref{fig:TOI_PS_real}).
   
\subsubsection{The thermal Sunyaev Zel'dovich signal}
 \label{sec:sz_simu}
       	\begin{table}
	\begin{center}
	\begin{tabular}{llccc}
	\hline
	\hline
	 & Compact cluster & Diffuse cluster \\
	\hline
	$\alpha$ & 1.2223 & 1.2223 \\
	$\beta$ & 5.4905 & 5.4905 \\
	$\gamma$ & 0.7736 & 0.7736 \\
	$r_{\mathrm{s}}$ (kpc) & 383 & 800 \\
	$\theta_{\mathrm{s}}$ (arcmin) & 1.1 & 2.3 \\
	$P_0 \ (\mathrm{keV/cm}^3)$ & 0.5 & 0.18 \\
	Best fit $\theta_{\mathrm{s}}$ (arcmin) & $1.048 \pm 0.042$ & $2.019 \pm 0.075$ \\   
	Best fit $P_0 \ (\mathrm{keV/cm}^3)$ & $0.449 \pm 0.052$ & $0.150 \pm 0.010$ \\  
	\hline
	\end{tabular}
	\end{center}
	\caption{Generalized Navarro, Frenk, and White parameters used to simulate the compact and diffuse clusters. The choice of the slope parameters is given in Sect.~\ref{sec:mcmc}. The last two lines provide the recovered marginalized best fit profile of the simulated clusters.}
	\label{tab:table_simu}
	\end{table}	
We use the generalized Navarro, Frenk, and White (gNFW) pressure profile \citep{gNFW} to describe the cluster pressure distribution out to a significant fraction of the virial radius. This profile is given by
\begin{equation}
	P(r) = \frac{P_0}{\left(\frac{r}{r_{\mathrm{s}}}\right)^{\gamma}\left(1+\left(\frac{r}{r_{\mathrm{s}}}\right)^{\alpha}\right)^{\frac{\beta-\gamma}{\alpha}}},
	\label{eq:gNFW}
   \end{equation}
where $P_0$ is a normalizing constant; $\alpha$, $\beta$, and $\gamma$ set the slope at intermediate, large, and small radii respectively, and $r_{\mathrm{s}}$ is the characteristic radius. The same profile can be written in its universal form \citep{arnaud_2010} by relating the characteristic radius to the concentration parameter $c_{500}$, $r_{\mathrm{s}} = r_{500}/c_{500}$, with $r_{500}$ the radius within which the mean density of the cluster is equal to 500 times the critical density of the Universe at the cluster redshift. The pressure normalization can then be written as $P_0 = P_{500} \times \mathds{P}_0$, where $\mathds{P}_0$ is a normalization factor and $P_{500}$ is the average pressure within the radius $r_{500}$ (related to the average mass within the same radius, $M_{500}$, by a scaling law). We finally define $\theta_{\mathrm{s}} = r_{\mathrm{s}}/D_A$ the characteristic angular size, where $D_A$ is the angular diameter distance of the cluster.

We simulate two different kinds of clusters. The first one is similar to what we observe for \mbox{RX~J1347.5-1145} in terms of amplitude and spatial extension (referred to as compact in the following). The second one is more diffuse, but its peak amplitude is the same as for the previous (referred to as diffuse hereafter). The corresponding parametrization can be found in Table~\ref{tab:table_simu}. Using these sets of parameters, we compute the Compton parameter map according to Eq.~(\ref{eq:y_compton}) by integrating the pressure along the line-of-sight. The map is then convolved with the instrumental beam and converted into surface brightness. We use the same scanning strategy, as in the case of \mbox{RX~J1347.5-1145} observations of {\it NIKA} during the Run 5. The focal plane and the number of detectors are also the same. The clusters are centered at the tSZ signal maximum decrement that we observe on \mbox{RX~J1347.5-1145}.

 \subsubsection{Validation of the pipeline on simulated data}
 \label{sec:valid_pipe_simu}
\begin{figure*}
	\centering
	\includegraphics[width=4.5cm]{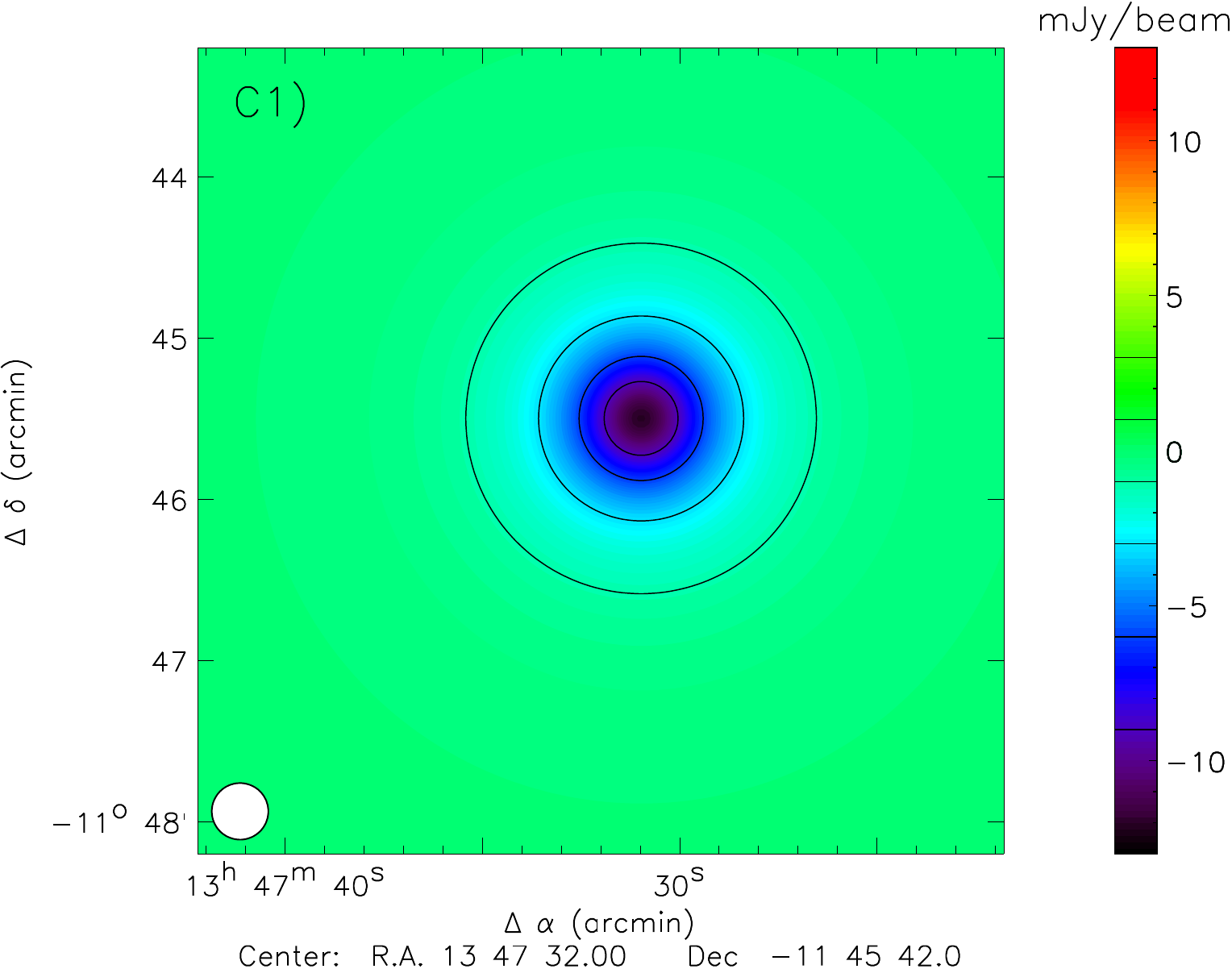}
	\includegraphics[width=4.5cm]{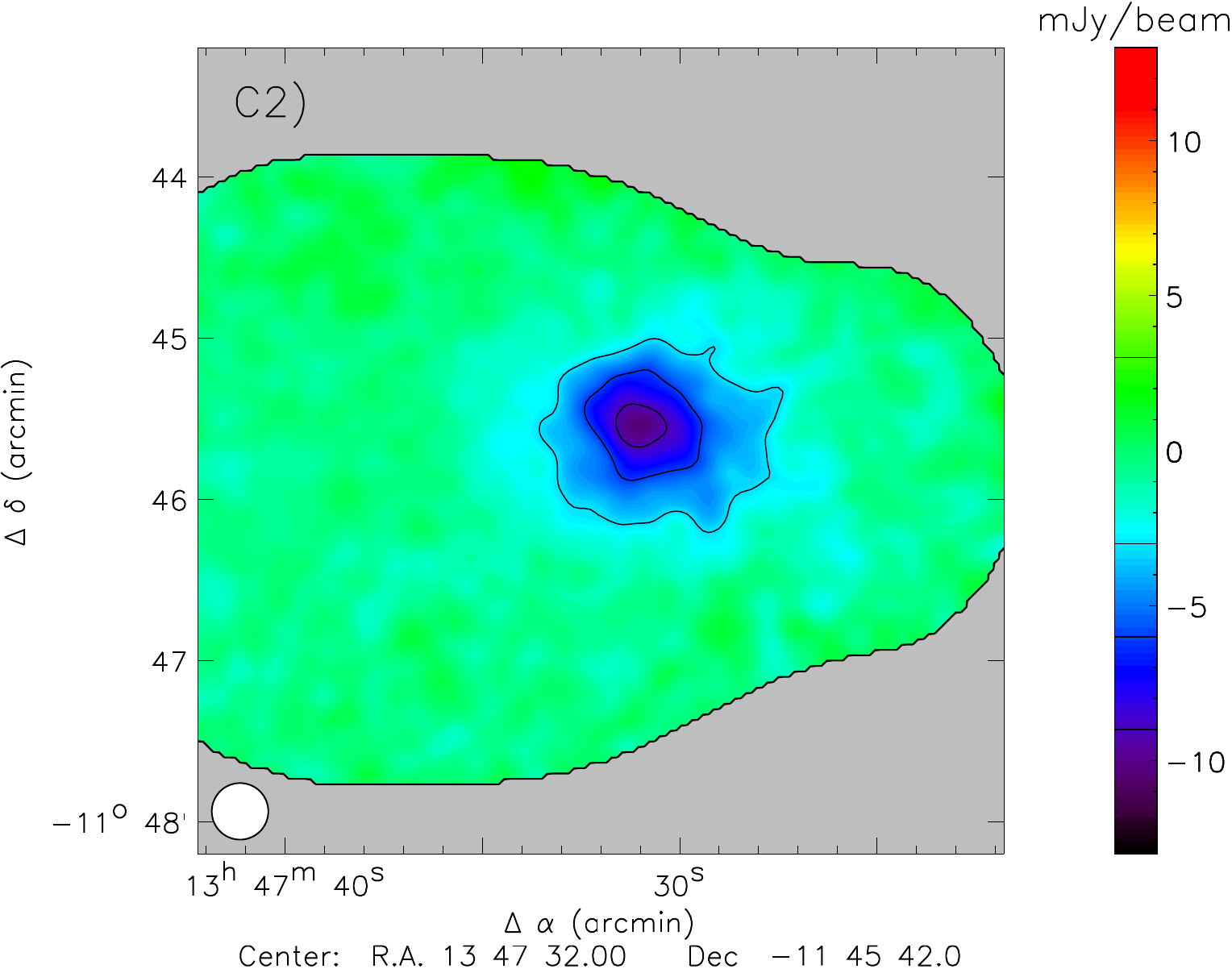}
	\includegraphics[width=4.5cm]{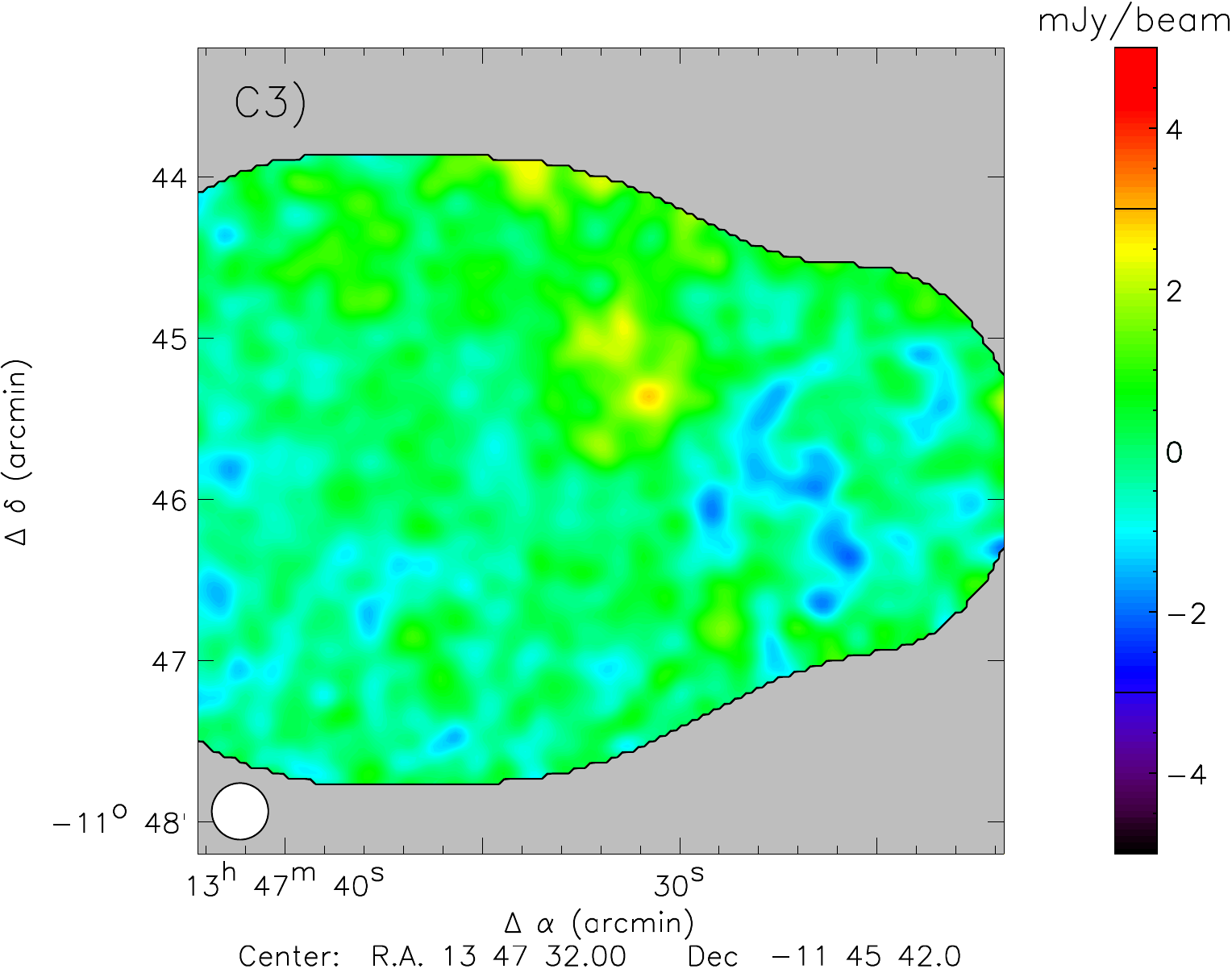}
	\includegraphics[width=4.5cm]{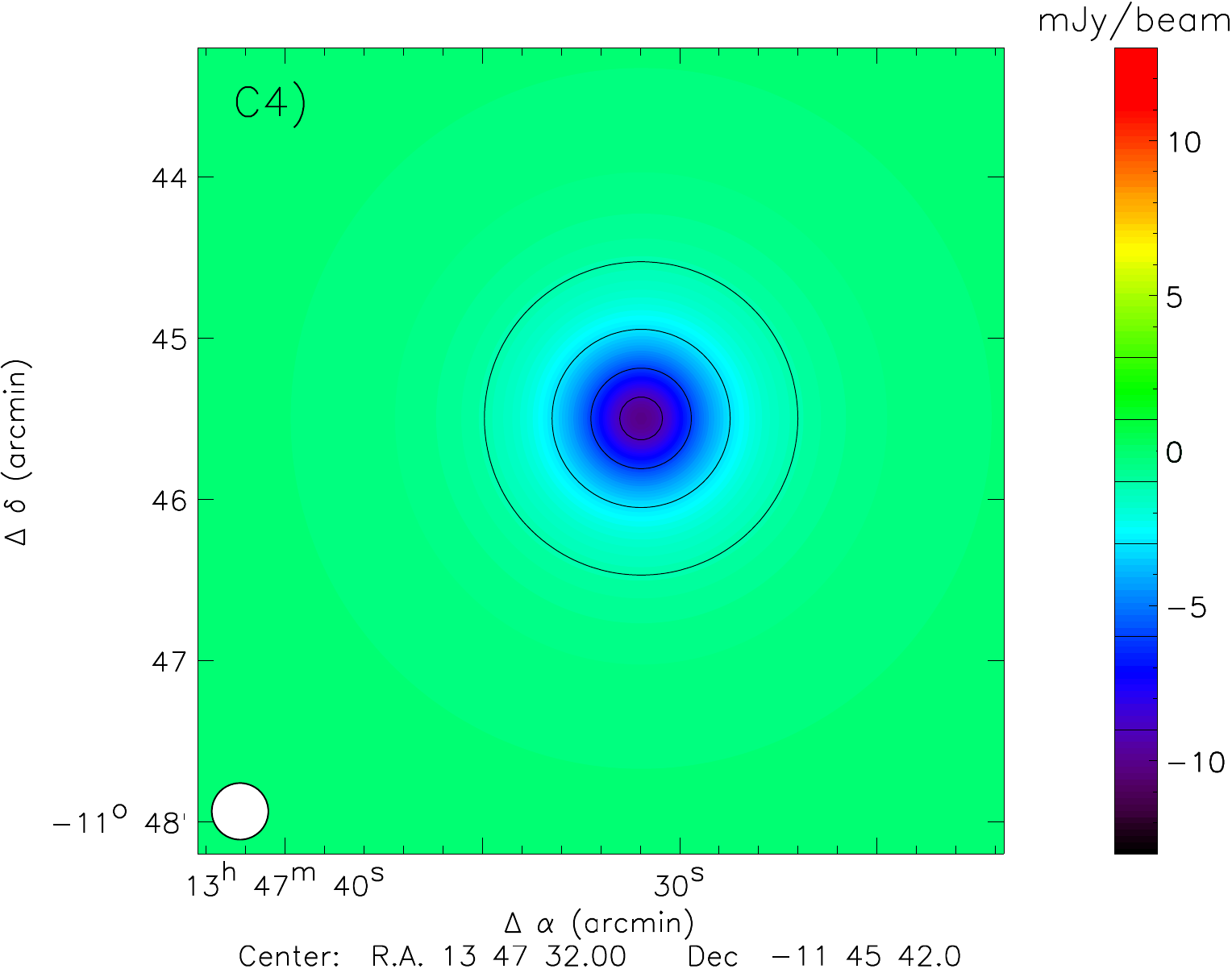}	
	\includegraphics[width=4.5cm]{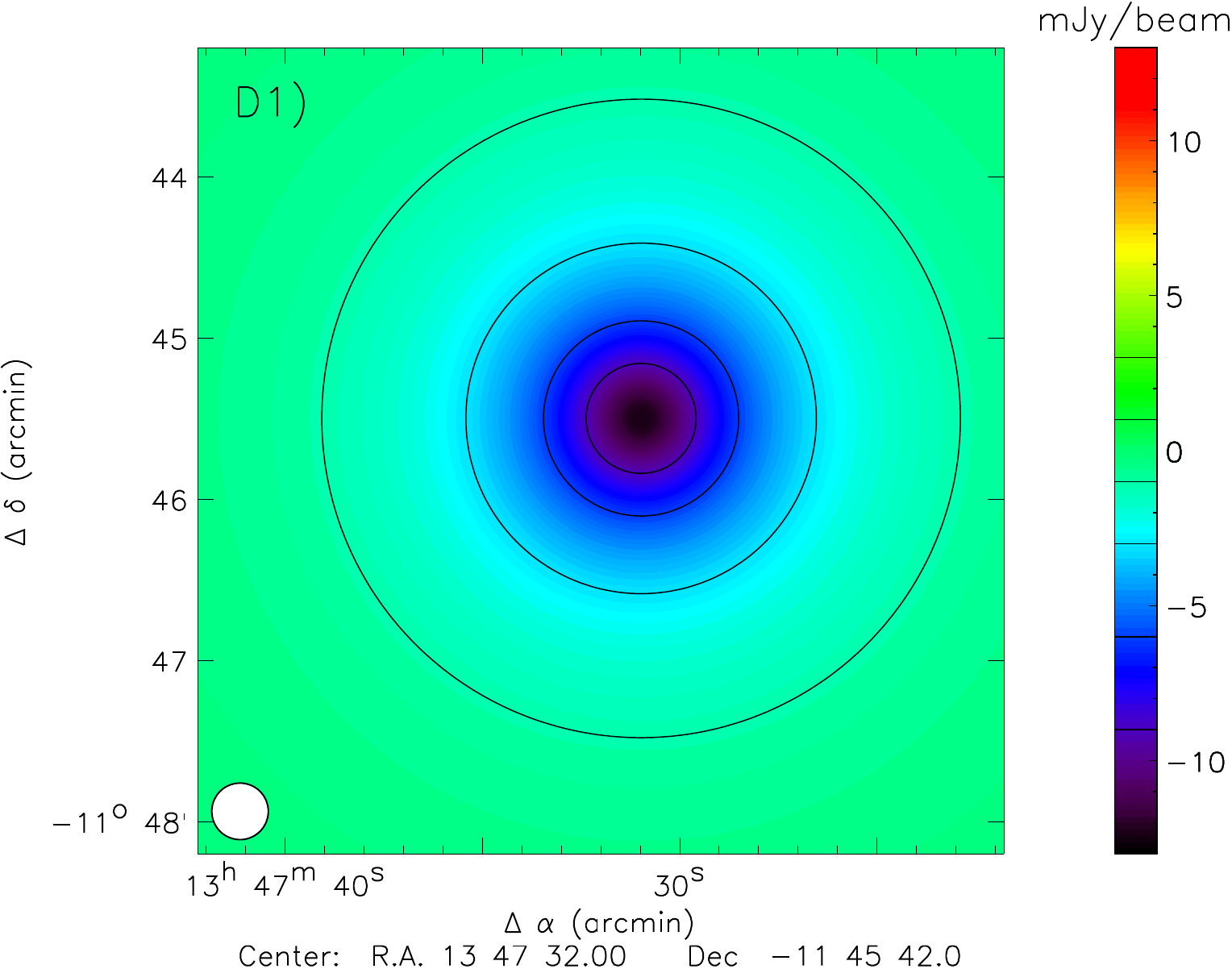}
	\includegraphics[width=4.5cm]{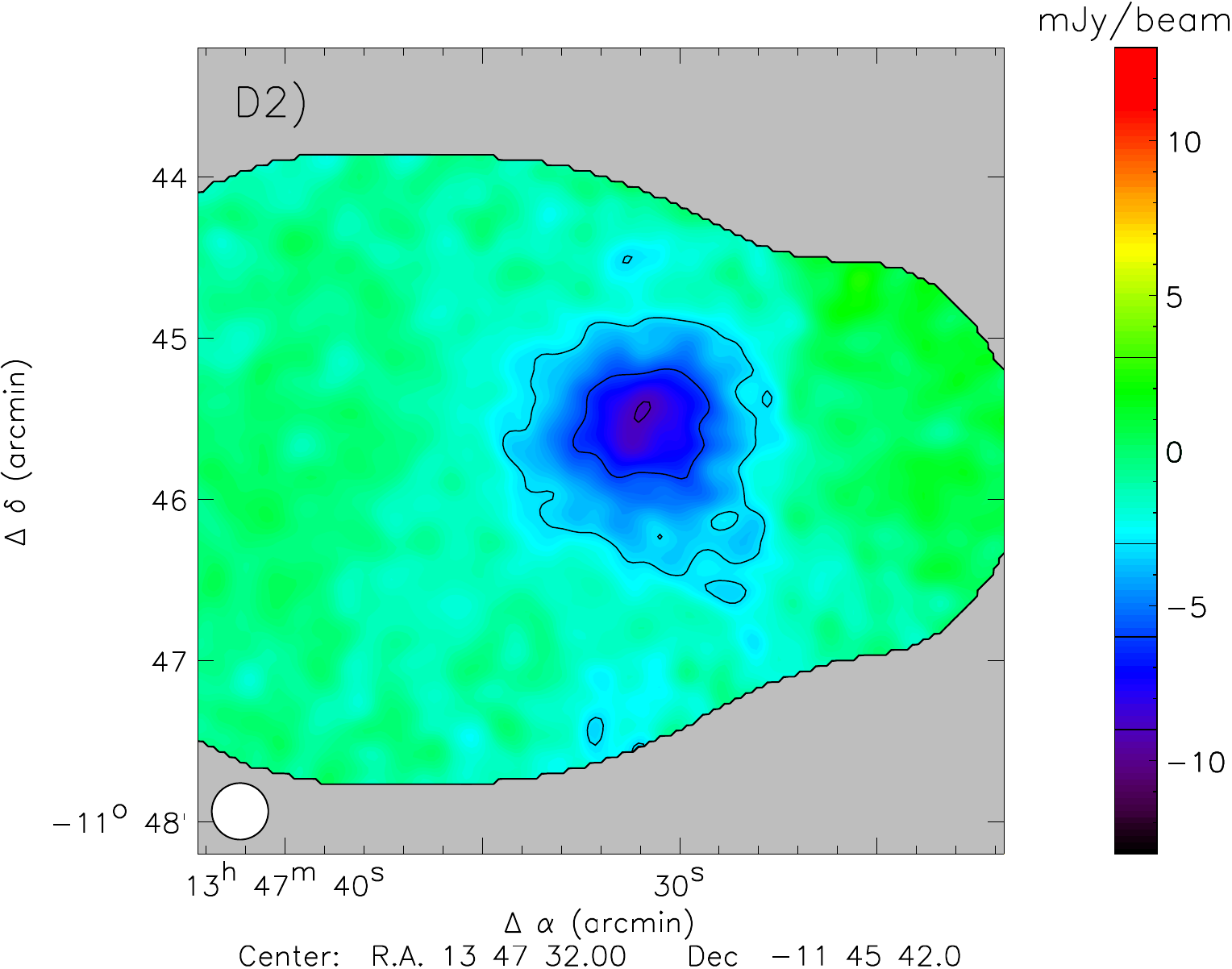}
	\includegraphics[width=4.5cm]{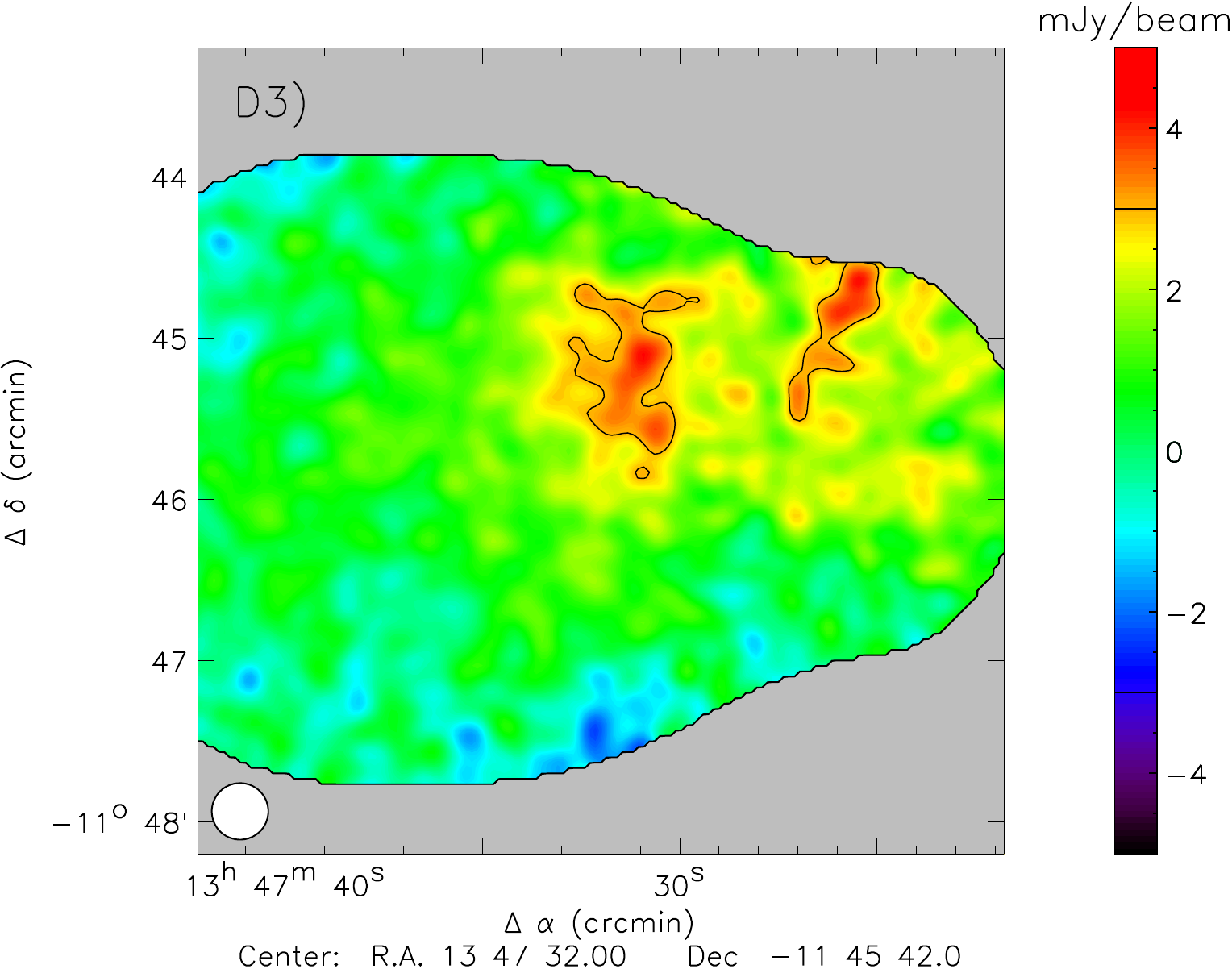}
	\includegraphics[width=4.5cm]{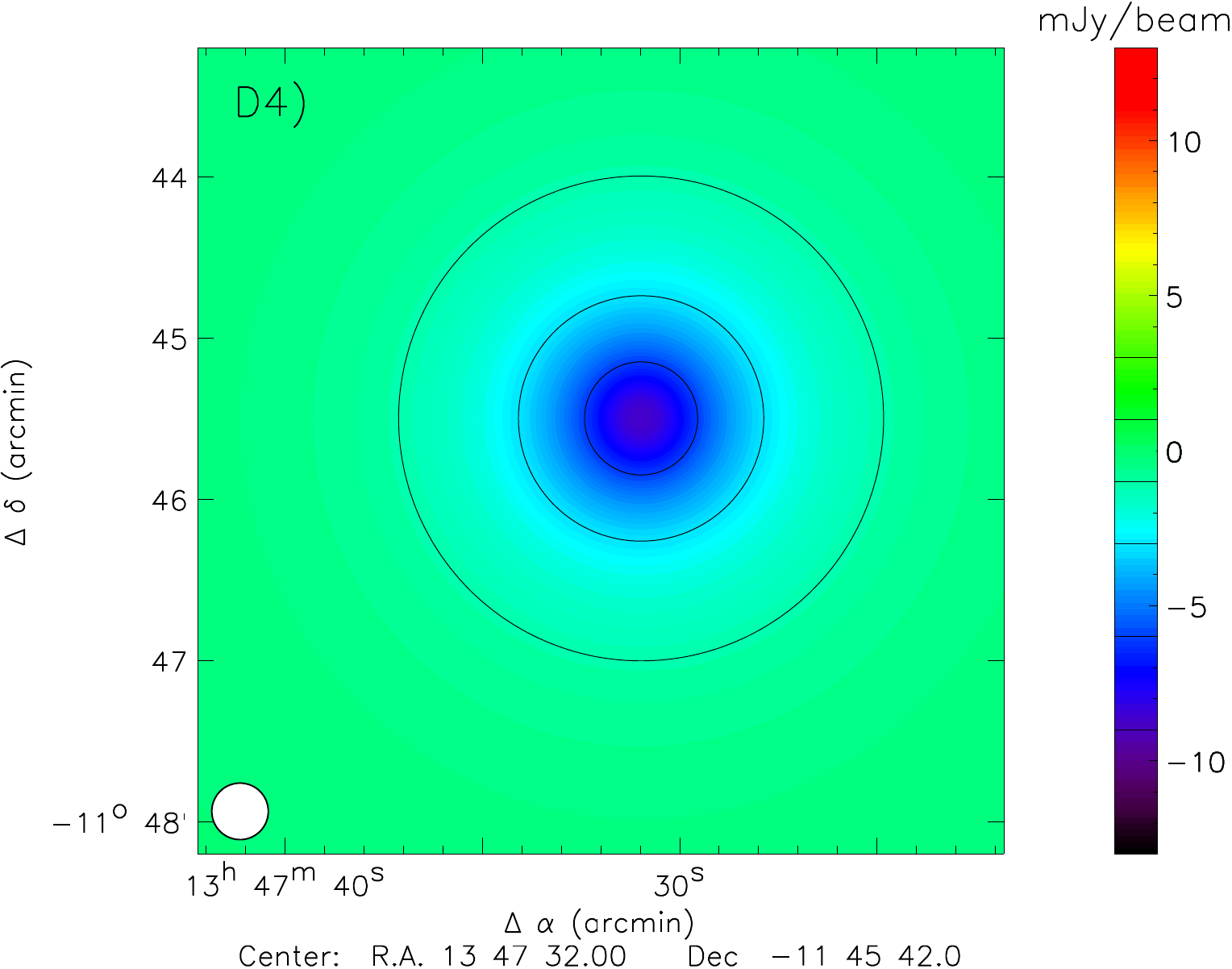}
	\caption{Generalized Navarro, Frenk, and White simulations of two clusters processed through the pipeline. The first one (compact cluster, as C) is similar to the {\it NIKA} map of \mbox{RX~J1347.5-1145} (top panels). The second (diffuse cluster, labeled D) is more extended (bottom panels). The parameters used in the cluster simulations are given in Table~\ref{tab:table_simu}. From left to right, we show the input model maps, the recovered maps, the residual maps, and the best fit model maps of the recovered signal. They are labeled from C1 to C4 and from D1 to D4 for the compact and diffuse cluster, respectively. The maps are shown up to a noise level that is twice the minimal noise level of the map. The effective beam is shown on the bottom left corner, accounting for the instrumental beam and an extra 10~arcsec Gaussian smoothing of the maps. The contours correspond to  3, -3, -6, and -9 mJy/beam, to which we add -1 mJy/beam for the model maps. The color scales range from -13 to 13 mJy/beam, except for the residual maps for which we have -5 to 5 mJy/beam. The center of the clusters has been simulated at the tSZ peak location of the {\it NIKA}  \mbox{RX~J1347.5-1145} map.}
        \label{fig:rxj_simu_map}
	\end{figure*}

After processing the simulated data, we recover the two cluster maps (compact is labeled C and diffuse is labeled D). Figure \ref{fig:rxj_simu_map} provides the input model maps, the recovered maps after the processing, the residual between the input models and the recovered maps, and the best fit models of the recovered maps. The top line corresponds to the compact cluster and the second to the diffuse cluster. The clusters are detected with a signal-to-noise of the order of 10 and are well mapped in both cases. The signal amplitude is slightly reduced with respect to the input maps.

\begin{figure}
	\centering
	\includegraphics[width=\columnwidth]{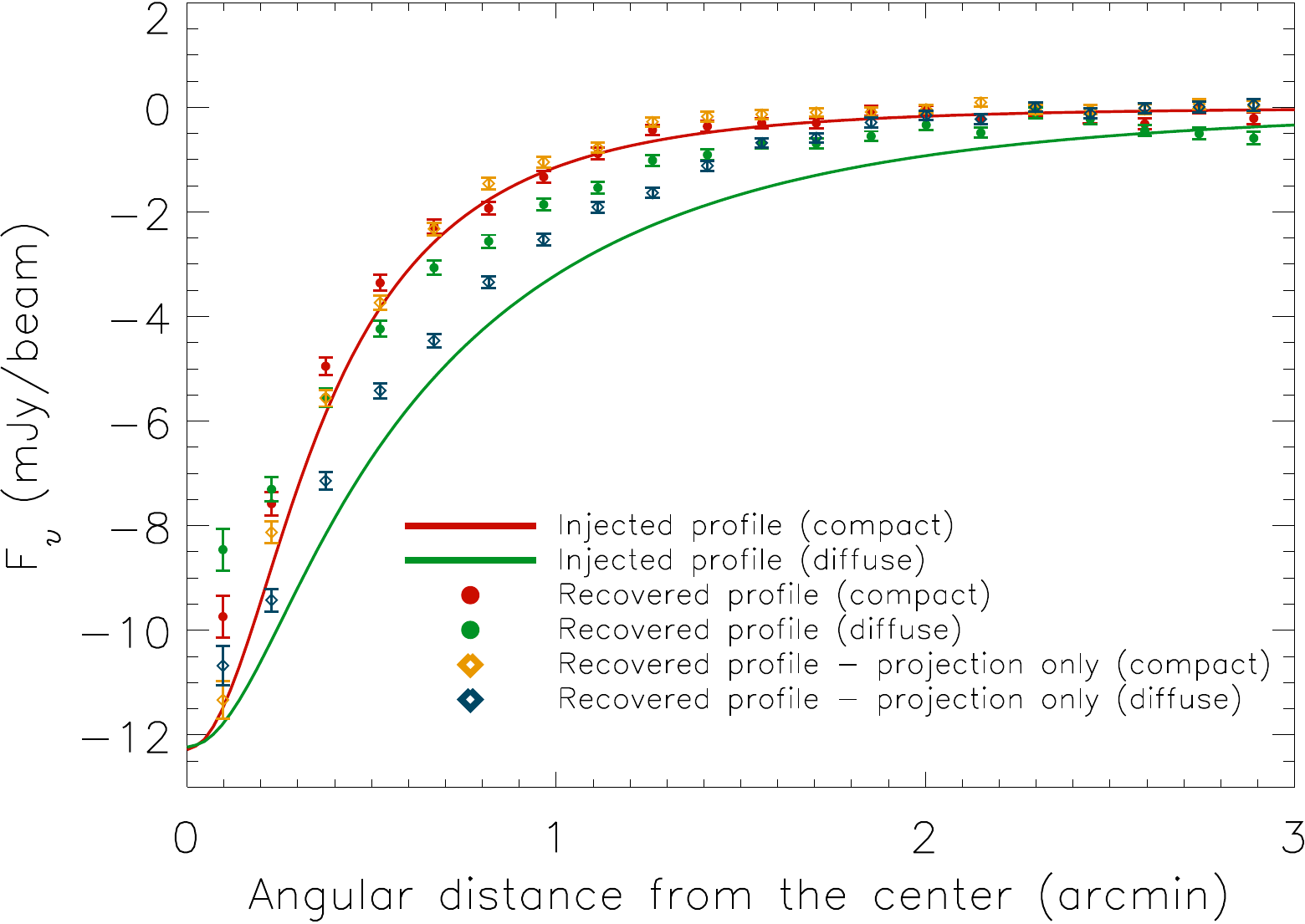}
	\caption{Comparison of the profiles injected in the simulation and recovered at the end of the pipeline. The injected profiles are given as red (compact cluster) and green (diffuse cluster) solid lines. The recovered profiles are shown with dots of similar colors. We also show the recovered profiles in the case of projection only without correlated noise, glitches, or pulse tube lines included in the simulation. They are given as orange (compact cluster) and blue (diffuse cluster) diamonds.}
        \label{fig:rxj_simu_prof}
	\end{figure}

Using these maps, we compute the angular profiles of the recovered clusters by evaluating the average flux value for a set of concentric annuli. They are shown in Fig.~\ref{fig:rxj_simu_prof}, as green and red dots for the compact and diffuse clusters, respectively. Comparison with the input profiles is provided by solid lines with similar colors. We also show the profiles recovered after projection only  to check zero-level effects (orange and blue diamonds for the compact and diffuse cluster, respectively), that is the input tSZ signal is simply projected without decorrelation or filtering.
	
First of all, due to the scanning strategy, the largest angular scale that can be recovered in the map is 6~arcmin, which corresponds to the size of the observed map. In Fig.~\ref{fig:rxj_simu_prof}, we show the profile of the diffuse cluster (projection only, as blue diamonds) that reaches the zero-level at 3~arcmin, which is unlike the injected profile that extends farther. Hence, the data processing affects the map in the case of the diffuse cluster by reducing the measured flux up to 25\% at a radii of $\sim~1$~arcmin. This can also be observed on the residual map of the diffuse cluster that is positive around the cluster peak. In the case of the compact cluster, the amplitude of the profile is not affected by more than 10\%, and the corresponding residual map is consistent with noise. Concerning the shape of the reconstructed signal with respect to the input one, we observe a flat transfer function for angular scales between  0.4 and 4~arcmin for the compact cluster case. Finally, the remaining correlated noise can slightly contaminate the profile, but it is not significant once averaged on concentric annuli.

We use the simulated maps to fit the normalization $P_0$ and the characteristic radius $r_{\mathrm{s}}$ of the pressure profile. This is done using Markov Chain Monte Carlo techniques that are further detailed in Sect.~\ref{sec:mcmc} (when applied on the \mbox{RX~J1347.5-1145} data). The recovered parameters can be compared to the input ones to estimate filtering effects and possible biases. Once marginalized, we find that the recovered parameters are within 1$\sigma$ of the inputs for both $P_0$ (10\%) and $r_{\mathrm{s}}$ (5\%) in the case of the compact cluster. For the diffuse cluster, we find that $P_0$ and $r_{\mathrm{s}}$ are underestimated by 2.7 (17\%) and 3.7 (12\%) $\sigma$, respectively. The MCMC best fit maps are given in panels C4 and D4 of Fig.~\ref{fig:rxj_simu_map}.

The effect of the radio point source subtraction (see Section~\ref{sec:point_source}) has also been checked via the simulations. To do so, a radio point source mimicking that, which is present in the \mbox{RX~J1347.5-1145} cluster has been added to the simulated data. It has then been removed during the processing by assuming a flux 3$\sigma$ lower than the injected one. The results change by less than 1$\sigma$ for both $P_0$ and $r_{\mathrm{s}}$ either for the diffuse or the compact cluster case.

 \subsection{Map of the undetected galaxy cluster IDCS~J1426.5+3508}
 \label{sec:undetec_source}
We have also observed \mbox{IDCS~J1426.5+3508}, a faint high redshift ($z=1.75$) cluster of galaxies. These observations correspond to 5 hrs 41 min of unflagged on-source data in atmospheric conditions, which are slightly poorer but comparable to those described in Table~\ref{tab:table_obs} for \mbox{RX~J1347.5-1145}. For \mbox{IDCS~J1426.5+3508}, the expected tSZ decrement is $\sim 0.25$~mJy/beam at 140~GHz with an angular size of $\sim$ 2~arcmin \citep{idcs}. We, therefore, do not expect a detection, since its flux is below the standard deviation of the expected noise at the cluster location by a factor of $\sim$ 5. \\

\begin{figure}
\centering
\includegraphics[width=\columnwidth]{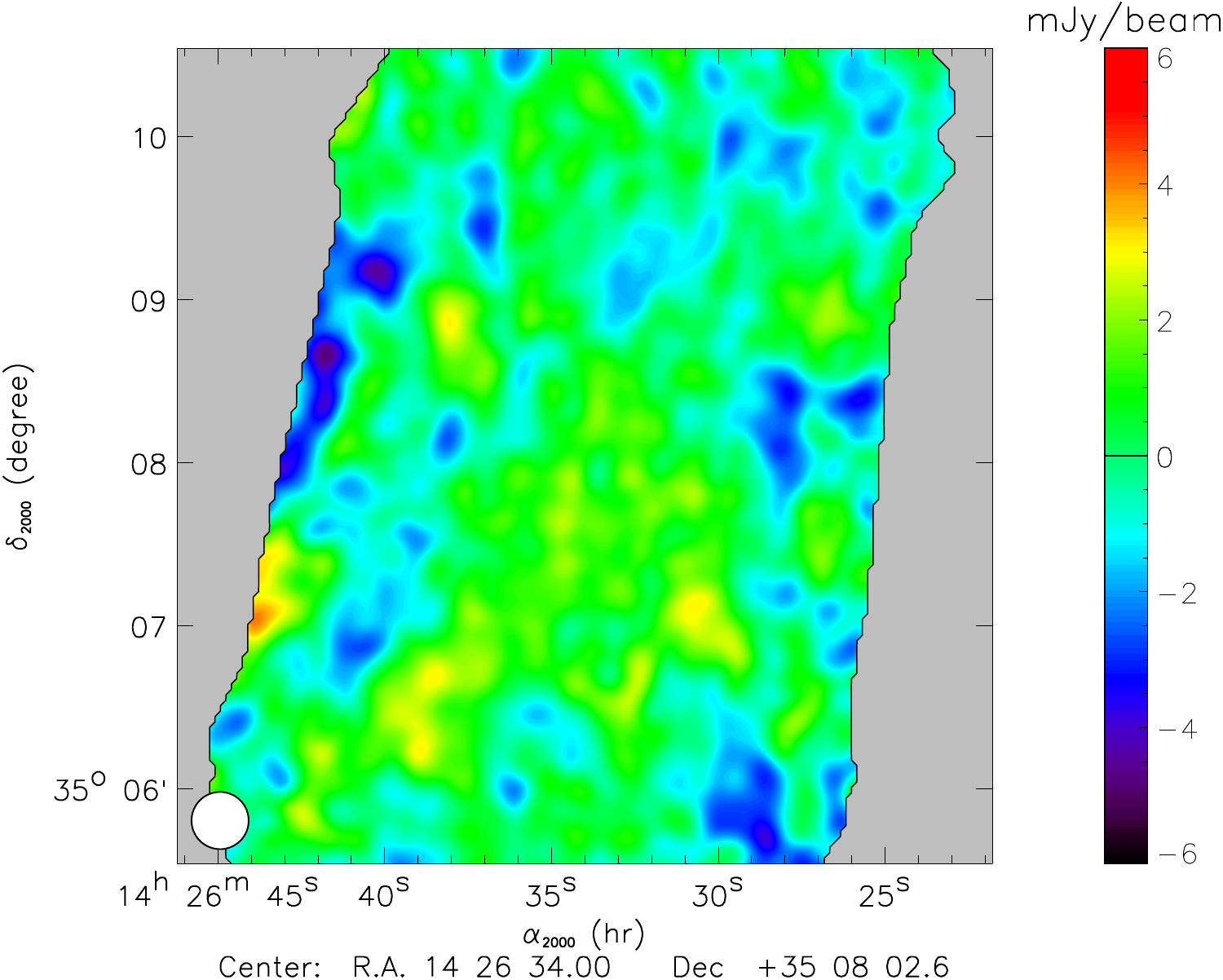}
\caption{Map at 140~GHz of the undetected galaxy cluster \mbox{IDCS~J1426.5+3508}.}
\label{fig:IDCS_J1426}
\end{figure}

In Fig.~\ref{fig:IDCS_J1426}, we show the map of \mbox{IDCS~J1426.5+3508} obtained after pipeline reduction. This map shows no evidence of tSZ signal, and it is consistent with noise as expected. 
This can be considered as a null test that allows us to conclude that the tSZ signal observed in the \mbox{RX~J1347.5-1145} data is not due to a bias in the analysis\footnote{We use \mbox{IDCS~J1426.5+3508} for a null test because we do not have observations of well-known empty fields that would better suit such a null test for the NIKA Run 5 campaign.}.

%% file: 06_result.tex
\subsection{\mbox{RX~J1347.5-1145} as observed by {\it NIKA}}
\label{sec:result_map}
\begin{figure*}
	\centering
	\includegraphics[width=0.45\textwidth]{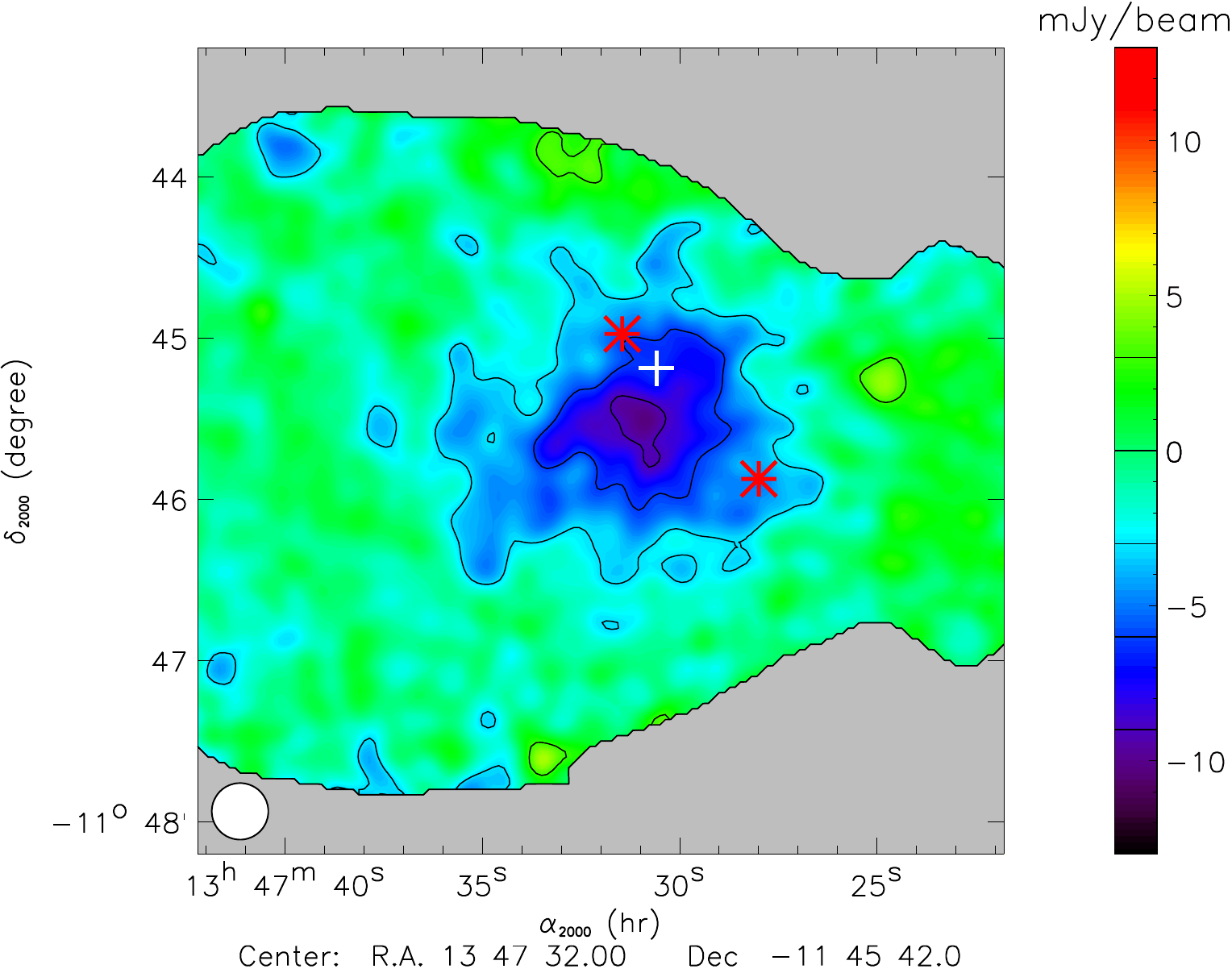}
	\hspace*{0.5cm}
	\includegraphics[width=0.45\textwidth]{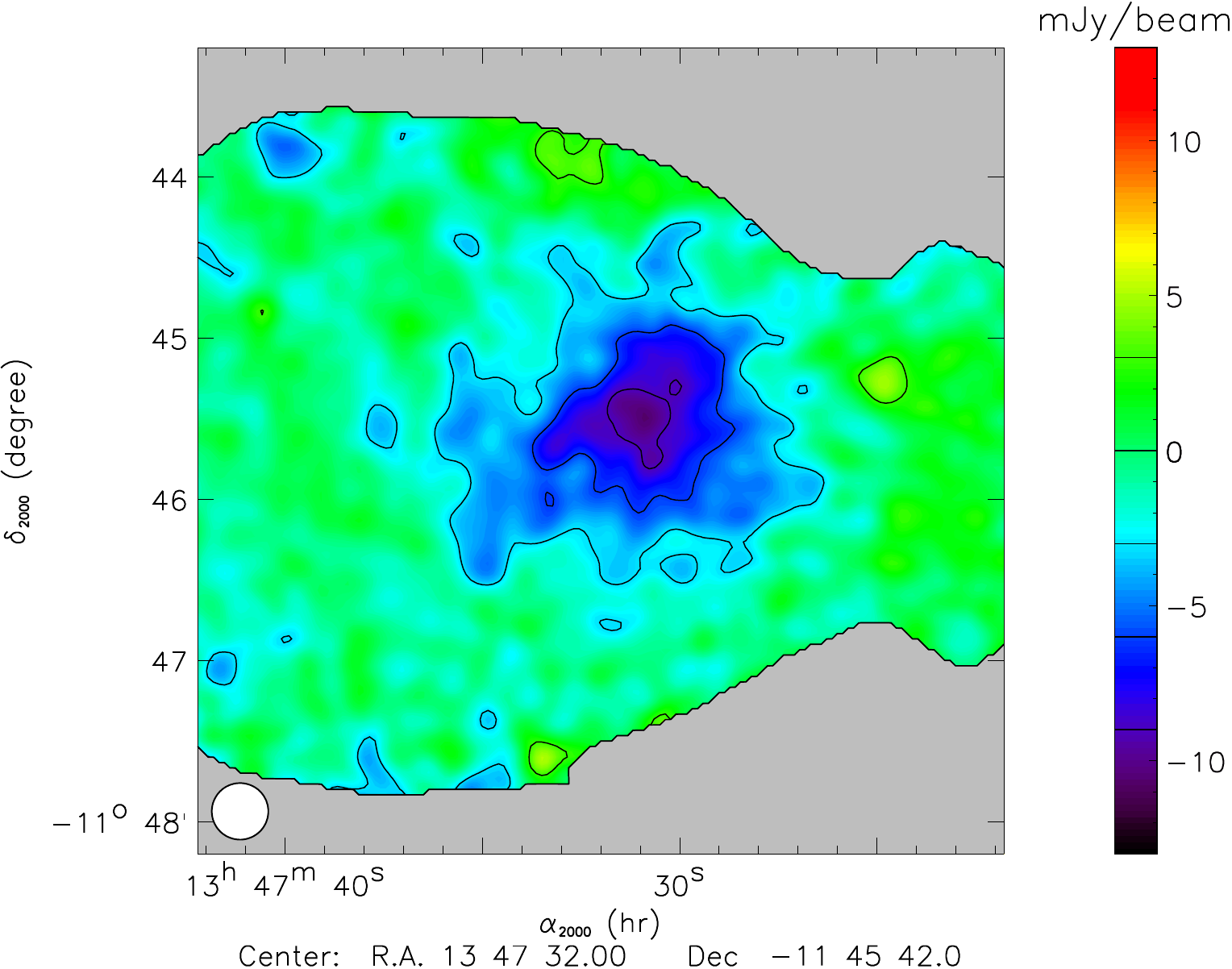}
	\caption{{\it NIKA} map of \mbox{RX~J1347.5-1145} at 140~GHz. Left: Original {\it NIKA} map with the radio source not subtracted. Right: Same map with the radio source subtracted. The maps are given in~mJy/beam. They are clipped up to a root mean square noise level that is twice the minimum of the map as detailed in the text. The contours are at 3, -3, -6 and -9~mJy/beam with $1 \sigma \cong 1$~mJy/beam at the cluster location. The minimum value of the maps corresponds to $y \simeq 10^{-3}$. The \mbox{X-ray} center location is represented by a white cross. The radio source location also corresponds to the white cross within 3~arcsec. The locations of the two infrared galaxies are given as red stars.}
        \label{fig:rxj}
	\end{figure*}

\begin{figure*}
	\centering
	\includegraphics[width=0.31\textwidth]{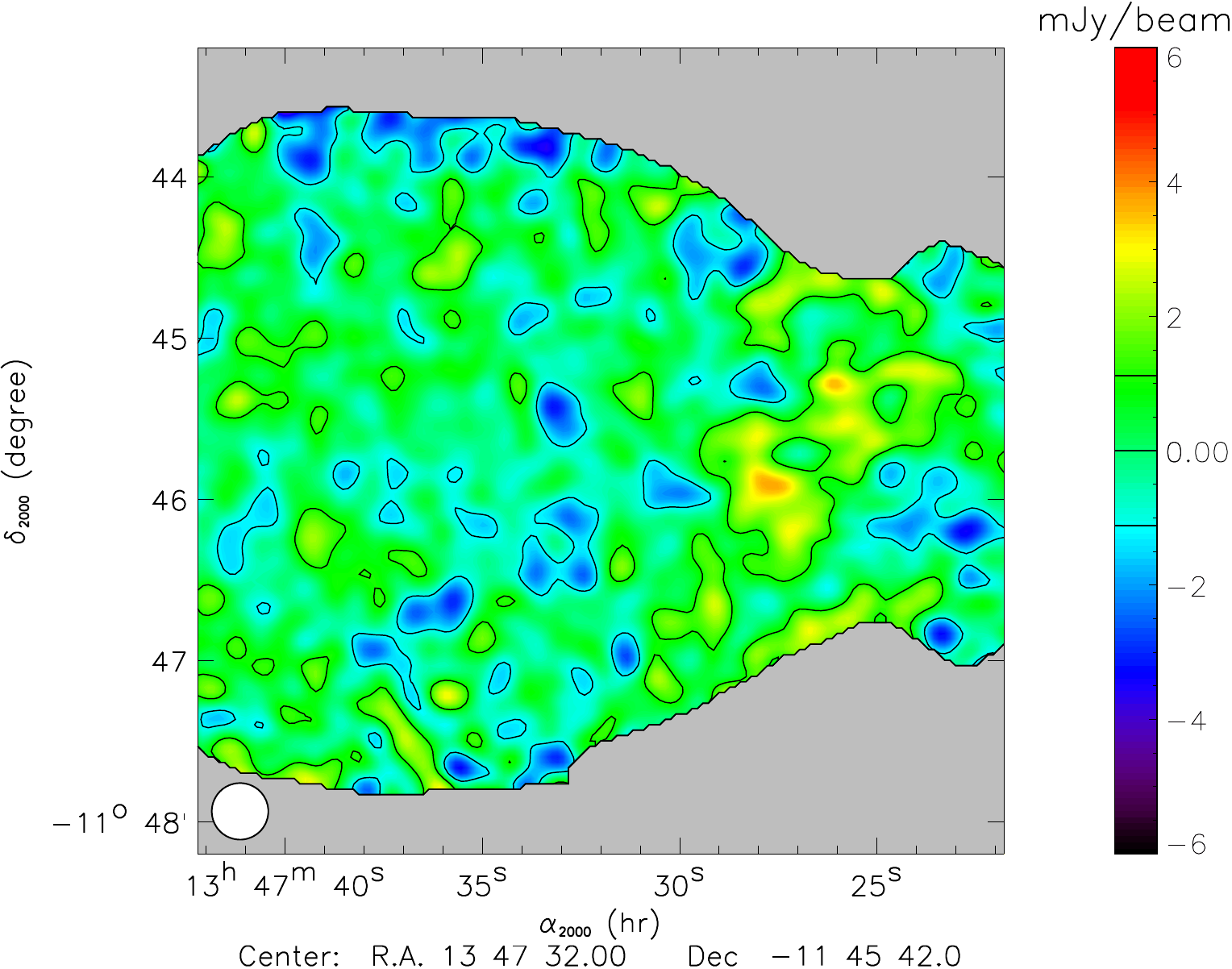}
	\hspace*{0.3cm}
	\includegraphics[width=0.31\textwidth]{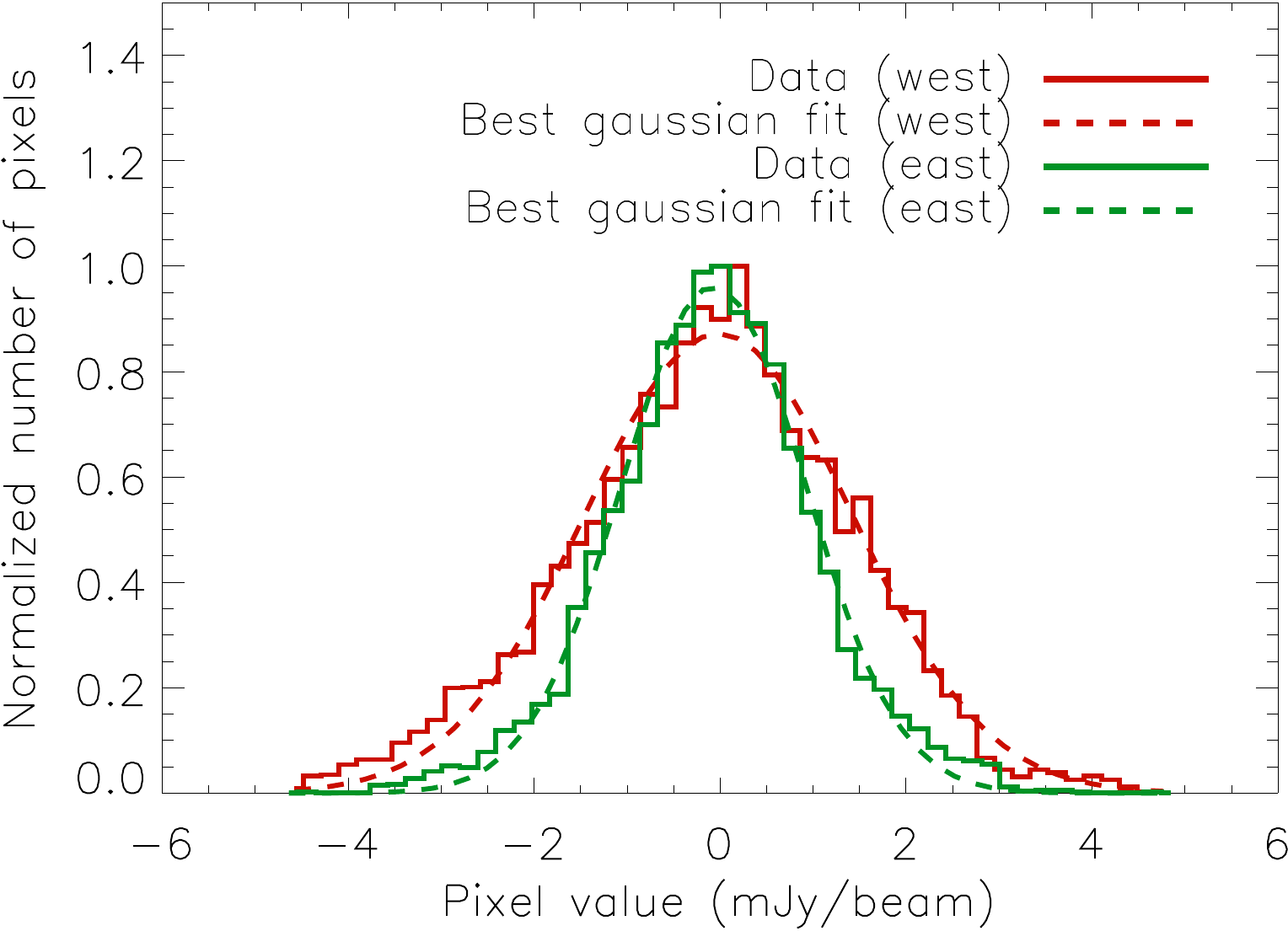}
	\hspace*{0.3cm}
	\includegraphics[width=0.31\textwidth]{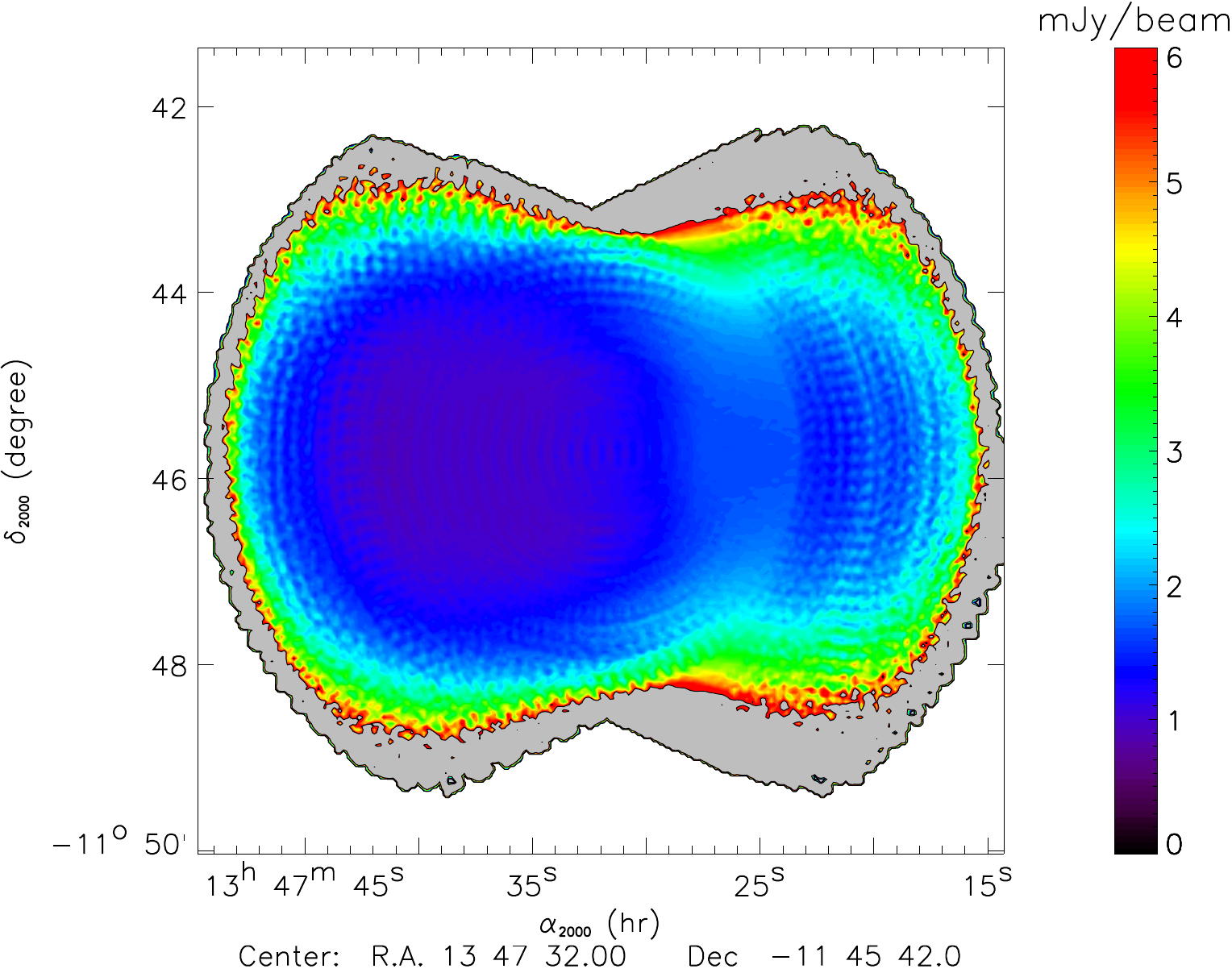}
	\caption{\mbox{RX~J1347.5-1145} observations. Left: Half-difference map of two equivalent subsamples mimicking the noise properties of the tSZ map. The pixels are 2$\times$2~arcsec, and the map has been smoothed with a 10~arcsec Gaussian filter, which is similar to the tSZ map of Fig.~\ref{fig:rxj}. The noise level is not homogeneous, which is lower on the left hand side, due to the differences of acquisition time. Middle: Noise distribution obtained from the half difference map. Since the noise is not homogeneous, we provide the distribution for both the eastern (left, green) and western (right, red) parts of the map. A Gaussian fit of the histograms gives the mean value of the standard deviation of the noise to be $<\sigma> = 0.99$~mJy/beam on the east side and $<\sigma> = 1.42$~mJy/beam on the west side. The minimum noise level reaches 0.8~mJy/beam. The contours of the noise map (left) correspond to the overall mean noise (i.e.  $\pm 1.11$ mJy/beam). Right: Standard deviation map estimated from difference maps. White regions have not been observed. Gray regions are those for which the standard deviation is higher than 6~mJy/beam.}
        \label{fig:rxj_jk}
	\end{figure*}

Figure~\ref{fig:rxj} presents the \mbox{RX~J1347.5-1145} tSZ map obtained with the {\it NIKA} prototype. The radio source is subtracted in the right panel but not in the left panel. The associated difference map of separated equivalent subsamples (Jack-Knife), which is normalized by a factor of 2 to preserve the statistical properties of the noise in the tSZ map, is given in the left panel of Fig.~\ref{fig:rxj_jk}. We also present on the middle panel of Figure~\ref{fig:rxj_jk} the histogram of the pixel values. The outside contour of the maps shown is defined by the limit where the statistical noise level, which increases toward the edges of the full map, equals twice the minimum noise level of the inner region. The bottle-like shape of the cut-off is due to the scan strategy detailed in Sect.~\ref{sec:obs_condition}. \\

The inhomogeneity of the noise can be seen directly on the half difference map in Fig.~\ref{fig:rxj_jk}. This is even more obvious on the histogram plot that provides the noise distribution in two different regions of the half difference map and on the standard deviation map. We observe that the standard deviation on the two regions is significantly different,  $<\sigma> = 0.99$~mJy/beam on the east side and $<\sigma> = 1.42$~mJy/beam on the west side. From the half difference map, we estimate the overall root mean square of the noise in the cluster map to $<\sigma> = 1.11$~mJy/beam. This is obtained by fitting the histogram of the pixel value with a Gaussian distribution. The contours overplotted on the tSZ maps of Fig.~\ref{fig:rxj} correspond to 3, -3, -6, and -9~mJy/beam with the noise level being $1 \sigma \cong 1$~mJy/beam at the cluster location. The beam is shown on the bottom left corner of the map, accounting for both the 18.5~arcsec instrumental beam and the extra 10~arcsec Gaussian smoothing of the map ({\it i.e.}, 21~arcsec). In terms of the Compton parameter, the sensitivity of the {\it NIKA} prototype camera during the campaign of November 2012 is $\sim 10^{-4} \sqrt{\mathrm{h}}$ for one beam and 1$\sigma$.
	
The maps in Fig.~\ref{fig:rxj} clearly show the tSZ decrement that reaches up to $\simeq 10 \ \sigma$. The signal is extended, and its maximum does not coincide with the \mbox{X-ray} center, (R.A,~Dec)~=~(13h~47m~30.59s,~-11$^{\mathrm{o}}$~45'~10.1"). It corresponds to the shock location, even for the radio point source subtracted map, which agrees with other single-dish observations.  As mentioned in Sect.~\ref{sec:previous_obs}, these results do not agree with those from {\it CARMA} interferometric \mbox{RX~J1347.5-1145}  observations \citep{plagge_2012}. The tSZ maximum corresponds to $\simeq 10^{-3}$ in units of Compton parameter $y$, as expected for this cluster according to \cite{pointecouteau_1999}. The consistency of the NIKA  \mbox{RX~J1347.5-1145} map with previous observations is further discussed in Sect.~\ref{sec:comparison}.
	 
\subsection{\mbox{RX~J1347.5-1145} profile}
\label{sec:profile}
\begin{figure*}
	\centering
	\includegraphics[width=0.45\textwidth]{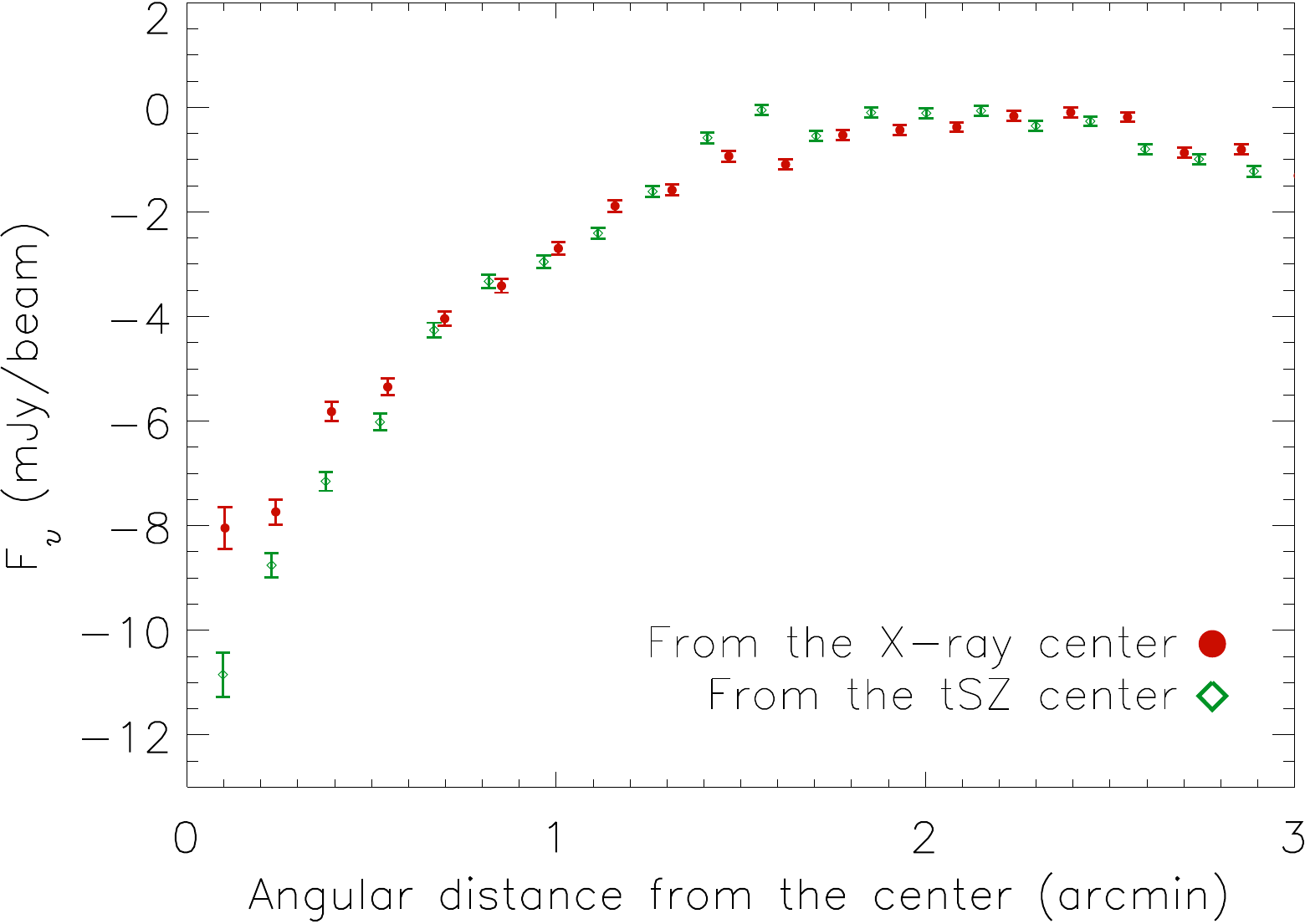}
	\hspace*{0.5cm}
	\includegraphics[width=0.45\textwidth]{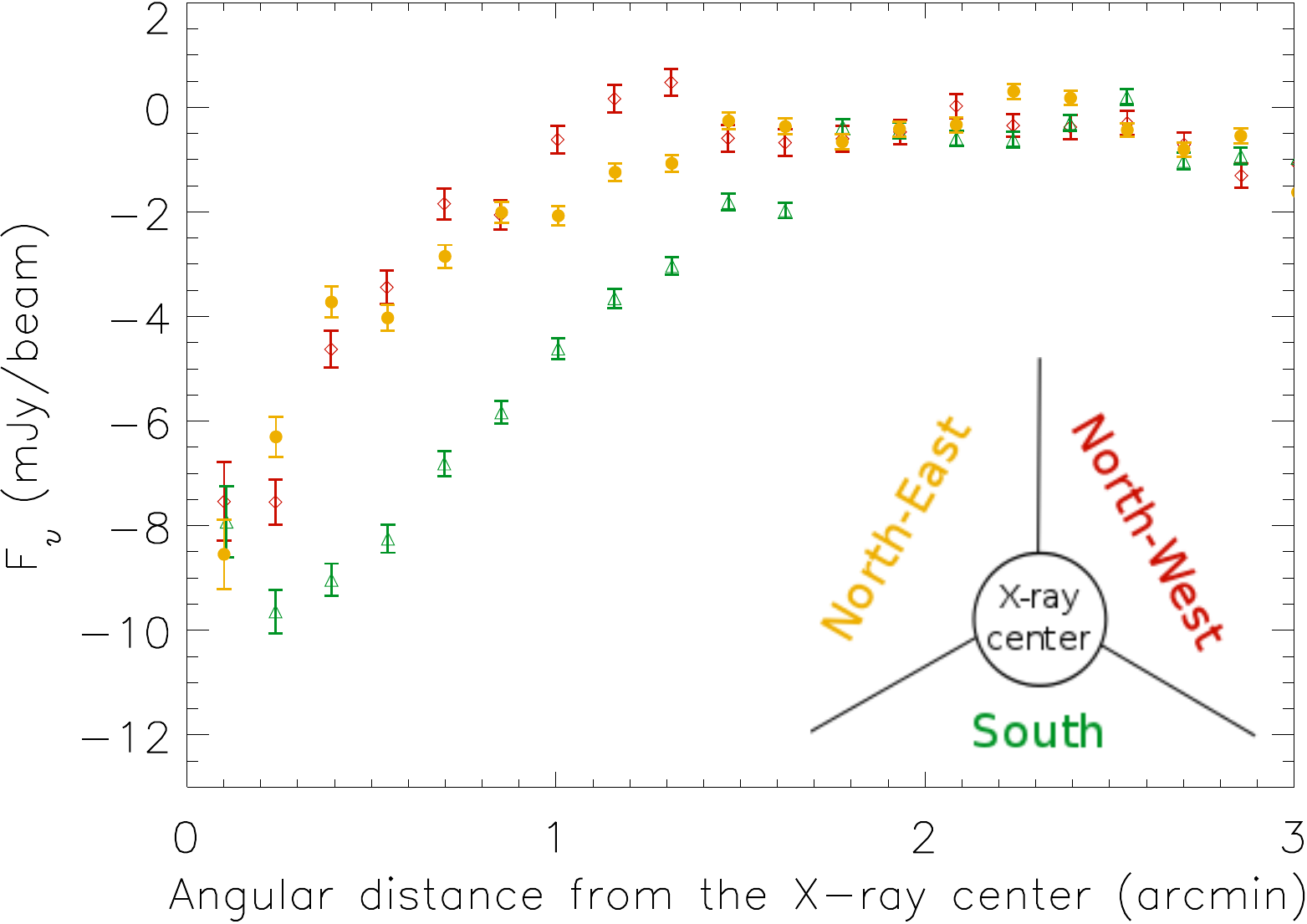}
	\caption{Radial flux profiles of \mbox{RX~J1347.5-1145}. Left: Comparison of the radial profile computed from the \mbox{X-ray} (red dots) and tSZ (green diamonds) centers. Right: Comparison of the radial flux profile in three different regions from the \mbox{X-ray} center. The map is cut from the \mbox{X-ray} center in three equal slices: one cut is vertical coming from the north to the center, and the two others are diagonal from the southeast and the southwest to the center, respectively. The red diamonds and yellow dots profiles correspond to the northwest and northeast part of the map, respectively, where the cluster is expected to be rather relaxed. The green triangle profile corresponds to the southern part of the map, where the merging occurred.}
        \label{fig:profiles}
	\end{figure*}

Figure~\ref{fig:profiles} gives the flux profile as a function of the angular distance that is extracted from the tSZ map in Fig.~\ref{fig:rxj}. In the case of \mbox{RX~J1347.5-1145}, the tSZ barycenter and the \mbox{X-ray} center do not coincide due to the ongoing merger. We compare the profile computed from the \mbox{X-ray} center, (R.A,~Dec)~=~(13h~47m~30.59s,~-11$^{\mathrm{o}}$~45'~10.1"), to the tSZ peak that is taken to be at the coordinates (R.A,~Dec)~=~(13h~47m~31s,~-11$^{\mathrm{o}}$~45'~30") from the maximum decrement of the {\it NIKA} map. The error bars have been computed from simulated noise maps with statistical properties estimated using the half-difference map presented on the left panel of Fig.~\ref{fig:rxj_jk}.

The right panel in Fig.~\ref{fig:profiles} compares the profile of \mbox{RX~J1347.5-1145} from the \mbox{X-ray} center in three different areas: the northwest, the northeast, and the south. It shows the increase in thermal pressure in the southern region, where the subclump (merging) is observed in \mbox{X-ray} and tSZ (Sect.~\ref{sec:previous_obs}). This is due to the compression of the hot gas within the merging process, which increases the temperature and thus the pressure (deepening the tSZ decrement at 140~GHz). We note that the southern extension coincides with the presence of a radio mini-halo \citep[see the work by][]{gitti_2007_bis}, which implies the presence of non-thermal electrons that could underline a non-thermal contribution to the total pressure (not seen in the tSZ signal). We also note that the radio source has been subtracted before the calculation of the profiles.

\subsection{Modeling of the cluster pressure profile}
\label{sec:mcmc}
The object \mbox{RX~J1347.5-1145} has been intensively studied in \mbox{X-rays}, which have revealed a fairly regular cluster at a large scale down to the center in the north direction with a low central entropy~\citep{Cavagnolo2009}. The contrast with the southern part, which exhibits a tSZ and \mbox{X-ray} extension, suggests that \mbox{RX~J1347.5-1145} was a spherical, relaxed cool-core cluster that is undergoing the merging of a subcluster on its southern part. We, therefore, aim at quantifying the tSZ South East extension detected with the {\it NIKA} prototype by modeling and subtracting the signal coming from the relaxed region, which is located on the northern-west side of the \mbox{X-ray} center. We model the tSZ signal by considering a gNFW profile (Eq.~\ref{eq:gNFW}), which is centered at the \mbox{X-ray} position of the system, whose inner, outer, and intermediate slopes ($\gamma$, $\beta$, $\alpha$) have been set equal to the cool-core best-fitting values of \cite{arnaud_2010} ($\gamma_\mathrm{cc} = 0.7736$, $\beta_\mathrm{cc} = 5.4905$, $\alpha_\mathrm{cc} = 1.2223$). The best-fitting values of $P_{0}$ and $\theta_s$ are obtained using a Markov Chain Monte Carlo (MCMC) approach. The sequence of random samples, known as the chain, has been built by implementing the Metropolis-Hasting algorithm \citep{Greenbert95}, which means that the parameter space is explored with a trial step drawn from a symmetric probability distribution. Convergence of the chains is checked by including the test proposed by \cite{GelmanRubin1992}.

	\begin{figure}	
	\centering
	\includegraphics[width=\columnwidth]{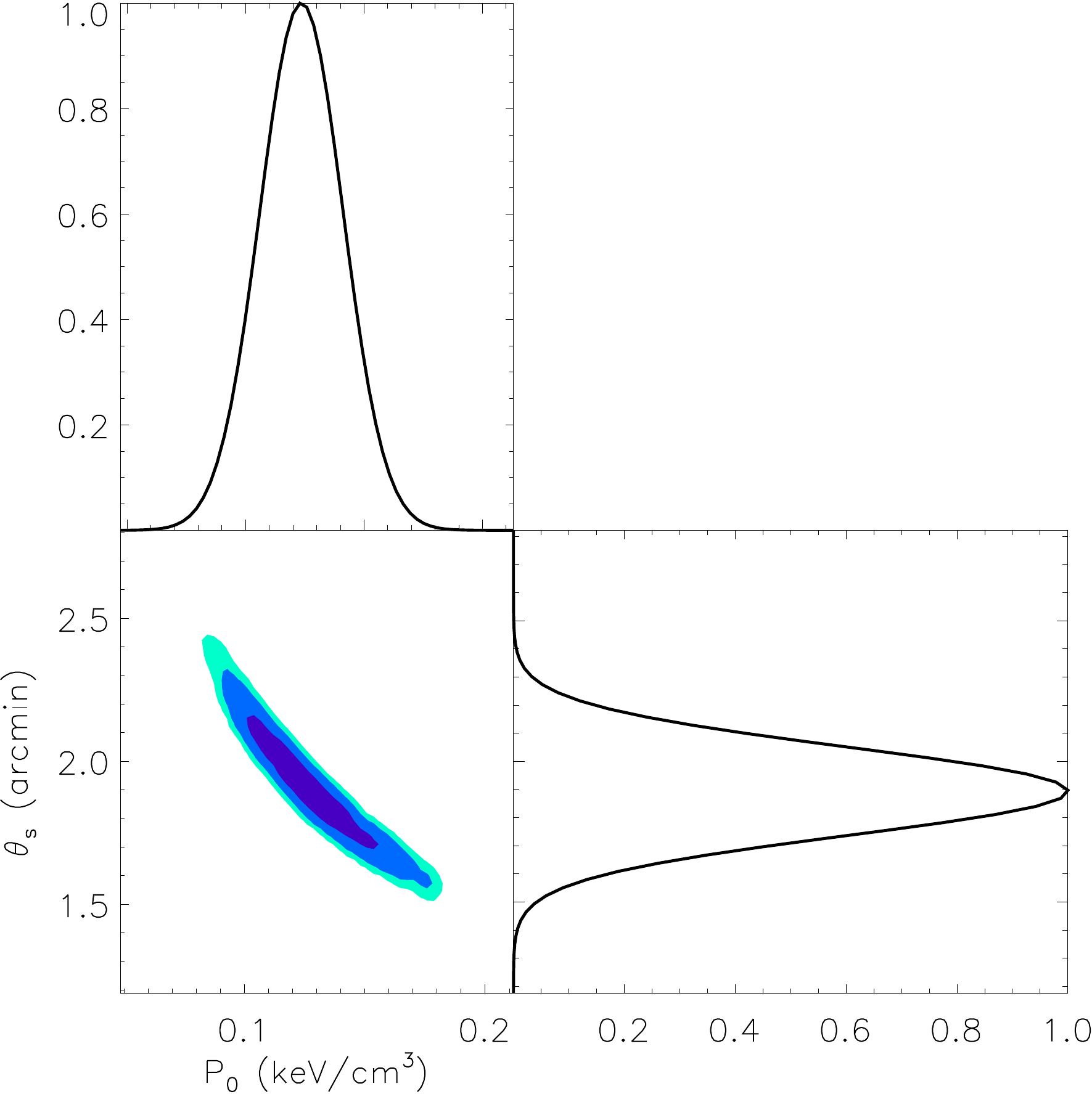}	
	\caption{Posterior likelihood of the MCMC pressure profile fit in the plane $P_0$ -- $\theta_s$. From dark to light blue, the colors correspond to 68\%, 95\%, and 99\% confidence levels. The top and right curves show the normalized Gaussian best fit of the marginalized likelihood of $P_0$ and $\theta_s$, respectively.}
        \label{fig:likelihood}
	\end{figure}

The parameters $P_{0}$ and $\theta_s$ have been constrained by masking the southeast extension. The mask has been defined as a half ring on the southern part of the cluster, centered on the \mbox{X-ray} peak with inner and outer radii set to 10 and 80~arcsec, respectively. By masking the hottest region of the system, the constraints obtained on the best fit parameters are mainly driven by the cool-core like component, where the cluster temperature remains below 10 keV. Consequently, the flux relativistic correction \citep{itoh_1998, nozawa1998, nozawa2006} is estimated to be $\lesssim$ 7\% at 140 GHz and needs to be propagated to the following results. The best fit parameters obtained are
    	\begin{eqnarray}
	P_{0} & = & 0.129 \pm 0.018 \ (\mathrm{stat.}) \ \pm^{0.035}_{0.025} \ (\mathrm{syst.}) \ \mathrm{keV/cm}^3 \  {\rm and}\nonumber \\
	\theta_s & = &  1.90 \pm 0.16 \ (\mathrm{stat.}) \ \pm^{0.38}_{0.00} \ (\mathrm{syst.}) \ \mathrm{arcmin}.  
	\label{eq:best_fit_nika}
	\end{eqnarray}
The corresponding posterior likelihood is given in Fig.~\ref{fig:likelihood} and accounts for statistical uncertainties only. The systematic uncertainties have been computed by using the calibration uncertainty and considering the bias filtering effect of the analysis that is estimated from the simulations described in Sect.~\ref{sec:sz_simu}. The pressure profile normalization parameter, $P_{0}$, is symmetrically affected by the calibration uncertainty, while the negative bias (lowering the true value) has been estimated to less than 20\%. The parameter $\theta_s$ is only affected by the bias, which is estimated to less than 20\% and lowers its true value. 

\begin{figure*}	
	\centering	
	\includegraphics[width=0.3\textwidth]{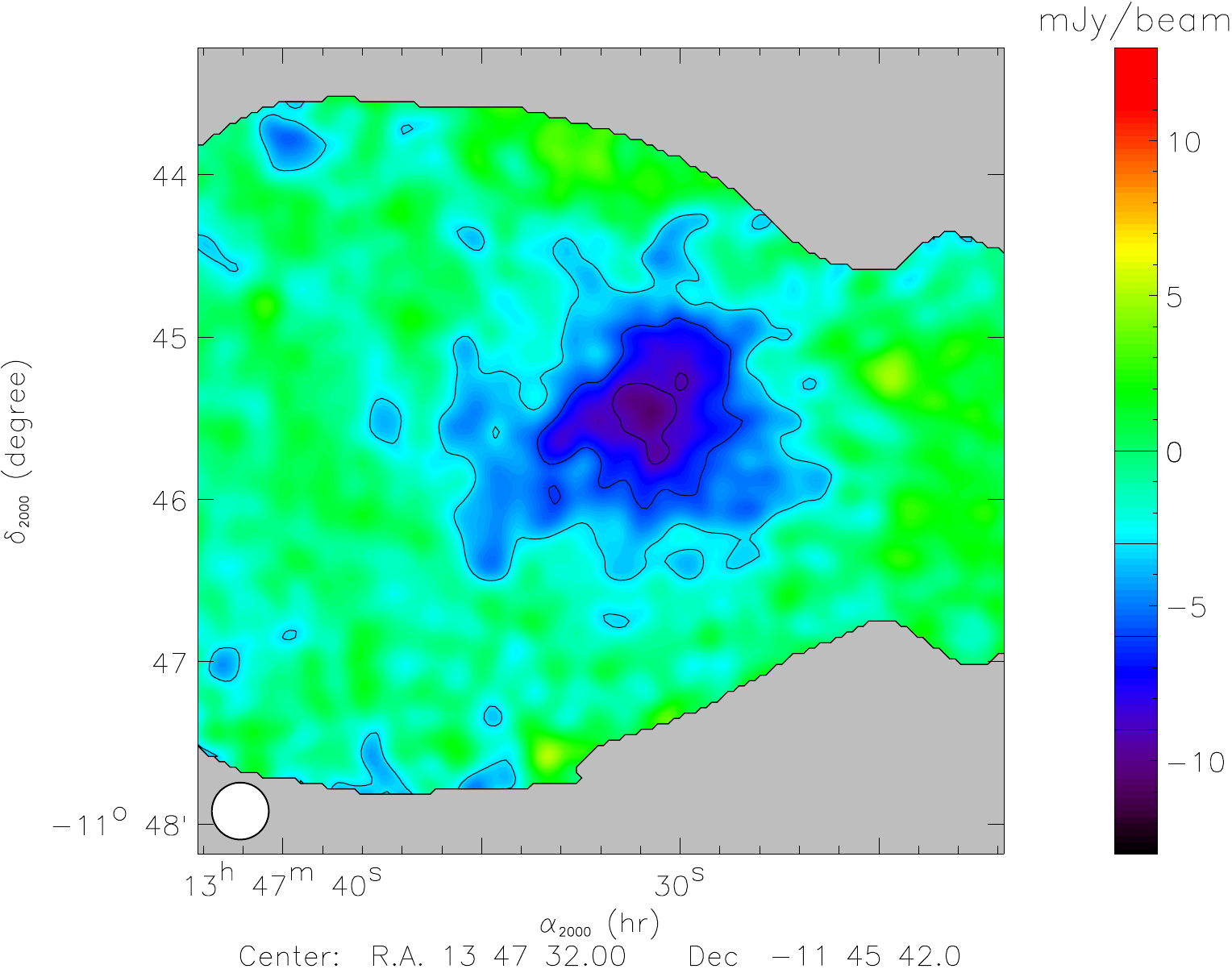}
	\hspace*{0.3cm}
	\includegraphics[width=0.3\textwidth]{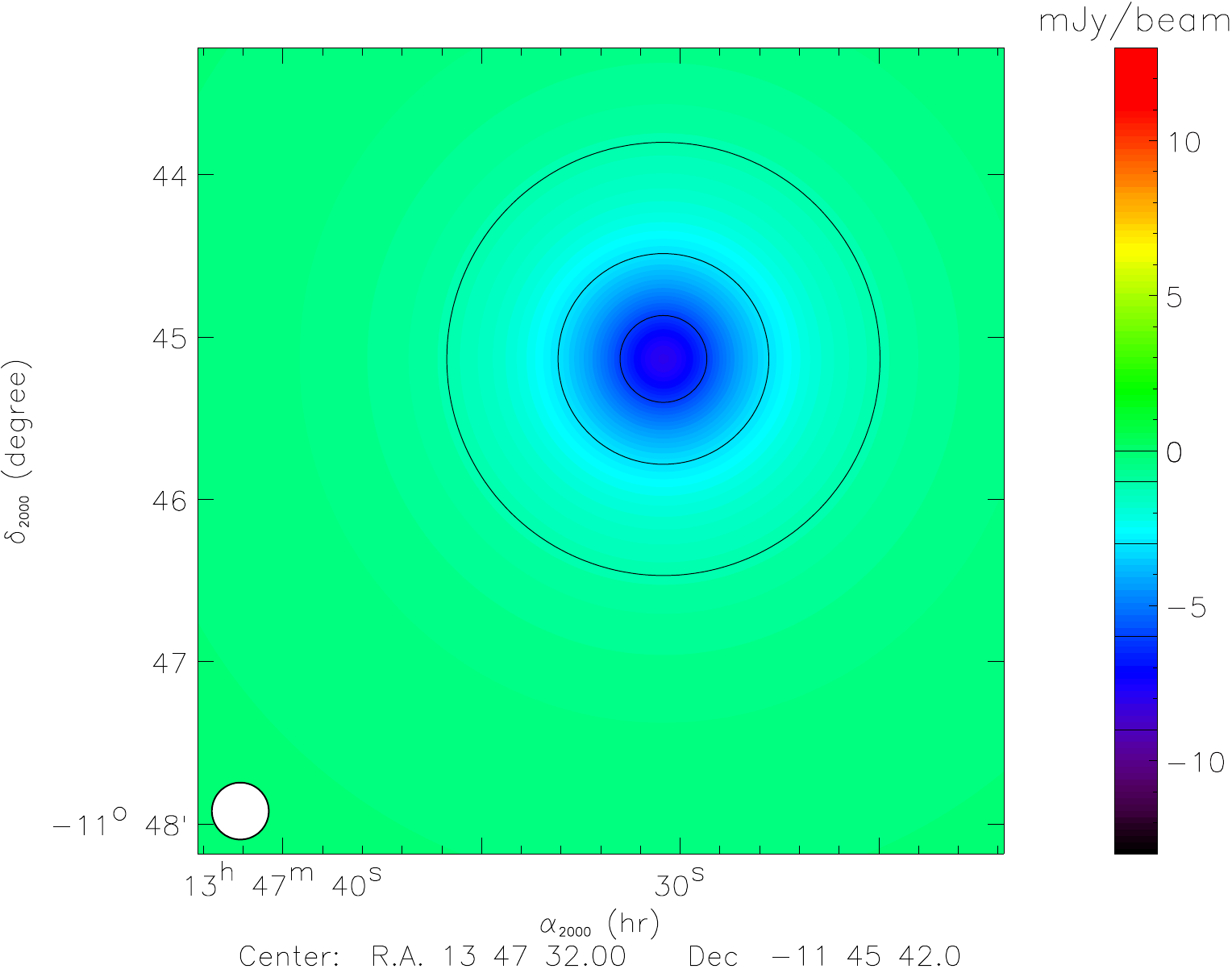}
	\hspace*{0.3cm}
	\includegraphics[width=0.3\textwidth]{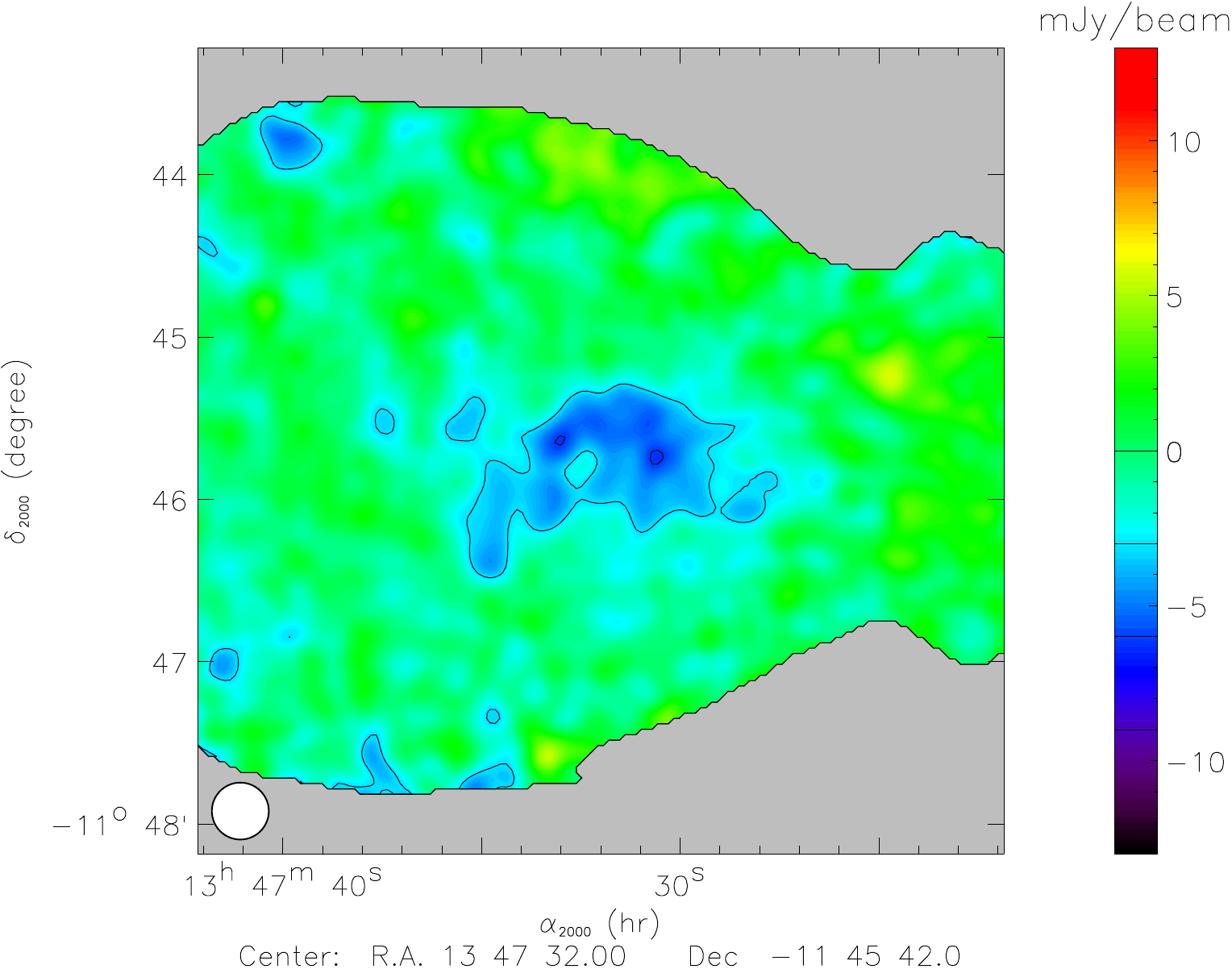}
	\caption{Comparison between the original point source subtracted \mbox{RX~J1347.5-1145} tSZ map (left panel) and the best fit model map excluding the shock area (middle panel). The residuals are given on the right panel map. The model accounts for the cluster emission well, except in the southern shocked area, as expected.}
        \label{fig:residual}
	\end{figure*}

Figure~\ref{fig:residual} compares the {\it NIKA} prototype point source subtracted map with the best fit model obtained for the relaxed component, and the residual. The model represents the northern part of the tSZ map well, but the southern side cannot be explained without including an overpressure component, which is known to be due to the merging of a subcluster (see Sect.~\ref{sec:previous_obs}).

The best fit model and the residual tSZ map, as given in Fig.~\ref{fig:residual}, have been used to quantify the distribution of the signal within the region, where the intracluster gas is more relaxed toward hydrostatic equilibrium, and the region, where it is expected to be shock heated. For this purpose, we compute the integrated Compton parameter, as defined as 
	\begin{equation}
	Y_{\theta_{\mathrm{max}}} =  \int_{\Omega(\theta_{\mathrm{max}})} y \ d\Omega,
	\label{eq:y_integ}
 	\end{equation}
	over the solid angle $\Omega$ up to the radius $\theta_{\mathrm{max}}$ from the \mbox{X-ray} center. This is separately done on the map and the residual (as seen in Fig.~\ref{fig:residual}). Given the size of the NIKA map, we integrate up to $\theta_{\mathrm{max}}$~=~2~arcmin. The total integrated Compton parameter within this radius is $Y^{\mathrm{total}}_{\theta_{\mathrm{max}}} =  (1.73 \pm 0.45) \times 10^{-3}$~arcmin$^2$. After removing the best fit cool-core model and integrating the residual in the same region, we obtain $Y^{\mathrm{shock}}_{\theta_{\mathrm{max}}} = (0.52 \pm 0.18) \times 10^{-3}$~arcmin$^2$. The errors on the integrated fluxes account for the statistical noise only. Systematic uncertainties are estimated to be of the order of 19\%. Thus, the shock contribution is estimated to be ($30 \pm 13 \pm 6$)~\% of the total tSZ flux at these small cluster-centric distances. Considering the {\it Planck} $Y_{5R_{500}}$ measurements \citep{Planck_fit}, the shock contribution corresponds to about 24 \% of the total tSZ flux. Previous observations by \cite{mason_2010} and \cite{plagge_2012} are consistent with a lower relative contribution of the shock of 9 to 10 \%. A direct comparison of these results with ours is difficult because of the very different methodologies used. In particular, the angular scales probed by the different instruments are not the same. Furthermore, \cite{mason_2010} and \cite{plagge_2012} have used external data to compute the overall tSZ flux, while we use {\it NIKA} data only in this paper.

%% file: 07_comparison.tex
\subsection{Comparison to the {\it Planck} catalog of tSZ sources}
The overall integrated Compton parameter for \mbox{RX~J1347.5-1145} can be compared to the {\it Planck} satellite measurement, as reported in the {\it Planck} catalog of tSZ sources \citep{Planck_survey}. For each detection, the {\it Planck} catalog provides the two-dimensional  $\theta_s$~--~$Y_{5r_{500}}$ probability distribution. The parameter $\theta_s$ is again the characteristic radius of Eq. \ref{eq:gNFW}, and $Y_{5r_{500}}$ is the integrated Compton parameter within a radius equal to 5$\times r_{500}$, therefore, assumed to be the total flux. The catalog also contains the slopes of the gNFW pressure profile used by the detection pipeline, which allows us to compute the $Y_{\theta_{\mathrm{max}}}$ / $Y_{5R_{500}}$ ratio, so to extrapolate the {\it Planck} flux ($Y_{5R_{500}}$) to the integrated signal at any cluster centric distance ($Y_{\theta_{\mathrm{max}}}$). To compare our result to {\it Planck} data, we have explored two different methodologies:

\begin{itemize}
\item {We fixed $\theta_s$ to its maximum likelihood value, obtaining $Y_{5R_{500}}~=~(2.17~\pm~0.36)~\times~10^{-3}$~arcmin$^2$ for \mbox{RX~J1347.5-1145}. Then, the $Y_{\theta_{\mathrm{max}}}$ / $Y_{5R_{500}}$ ratio returns $Y^{Planck}_{\theta_{\mathrm{max}}}$~=~$(1.78 \pm 0.30) \times 10^{-3}$~arcmin$^2$ for $\theta_{\mathrm{max}} = 2$~arcmin, which agrees with the {\it NIKA} value, $Y^{total}_{\theta_{\mathrm{max}}}$~=~$(1.73~\pm~0.45)~\times~10^{-3}$~arcmin$^2$.  
\item The {\it Planck} tSZ angular size ($\theta_s$) -- flux ($Y_{5r_{500}}$) degeneracy can also be broken by fixing $r_{500}$ to its \mbox{X-ray} derived value without changing the other pressure profile parameters. This includes $c_{500}$ (which is kept equal to 1.1733; the value given by \citealp{arnaud_2010}). This is what \citealp{Planck_survey} uses, when recovering the integrated tSZ signal within the \mbox{X-ray} size. Following this approach, we obtain $Y^{Planck}_{\theta_{\mathrm{max}}}$~=~$(1.23~\pm~0.21)~\times~10^{-3}$~arcmin$^2$ for $\theta_{\mathrm{max}}~=~2$~arcmin. This value is still consistent with the {\it NIKA} flux, although weaker. The latter can be understood if we consider that the \mbox{X-ray} derived $r_{500}$~=~1.42~Mpc~$\equiv~\theta_{500}$=~3.94~arcmin \citep[{\it MCXC},][]{MCXC} is much larger than the reliable radial extent of the {\it NIKA} map. The $r_{500}$ that can be deduced from the {\it NIKA} $\theta_s$ parameter is, by contrast, smaller. However, we expect the two integrated Compton parameters to converge when we move to larger $\theta_{\mathrm{max}}$. Indeed, when pushing the integration up to $\theta_{\mathrm{max}} = 2.5$~arcmin (by extrapolating the best-fit model of the relaxed region to angular distances not directly probed through our observations and assuming that the contribution due to the shock is negligible at scales larger than $\sim$ 2~arcmin), we obtain $Y^{Planck}_{\theta_{\mathrm{max}}}$~=~$(1.52~\pm~0.26)~\times~10^{-3}$~arcmin$^2$ versus $(1.77~\pm~0.45)~\times~10^{-3}$~arcmin$^2$ for {\it NIKA}. Alternatively, the larger {\it NIKA} flux can also be explained by relaxing the hypothesis of a constant $c_{500}$. The different sensitivities of {\it Planck}+{\it MCXC} and {\it NIKA} to the signal distribution could lead to differences in the recovered flux distribution. With $r_{500}$ fixed to its \mbox{X-ray} value, a larger value of $c_{500}$ implies a smaller value of  $\theta_s$ and, therefore, that a larger fraction of the total tSZ flux is located within the innermost regions. In this case, we would have a better consistency between  the {\it Planck}+{\it MCXC} and the {\it NIKA} integrated Compton parameter, even at smaller cluster-centric distances. Since the value of $c_{500}=1.1733$ has been obtained on an average (universal) profile~\citep{arnaud_2010} and \mbox{RX~J1347.5-1145} is known to have a very peaked morphology compared to other clusters, this hypothesis is likely to be correct: the inner slope parameter $\gamma$ being fixed, $c_{500}$ can typically vary from $\sim$ 0 to $\sim$ 5 between clusters~\citep{planck_pp}.}
\end{itemize}

From the comparison to the {\it Planck} data, we can conclude that {\it NIKA} is able to recover most of the tSZ signal, despite the large angular scale cutoff (above 3~arcmin). This is consistent with what was found in the simulations in Sect.~\ref{sec:valid_pipe_simu} and in particular for the compact cluster case, which is very similar to the {\it NIKA} \mbox{RX~J1347.5-1145} observations regarding the tSZ flux and angular extension. From this, we can convey that {\it Planck} and {\it NIKA} are complementary. This will be even more interesting and easily exploitable with the larger field of view (6.5~arcmin) that the {\it NIKA2} camera will reach.

\subsection{Comparison to {\it DIABOLO} tSZ observations}
\label{sec:comp_sz}
\begin{figure*}
	\centering
	\includegraphics[width=0.45\textwidth]{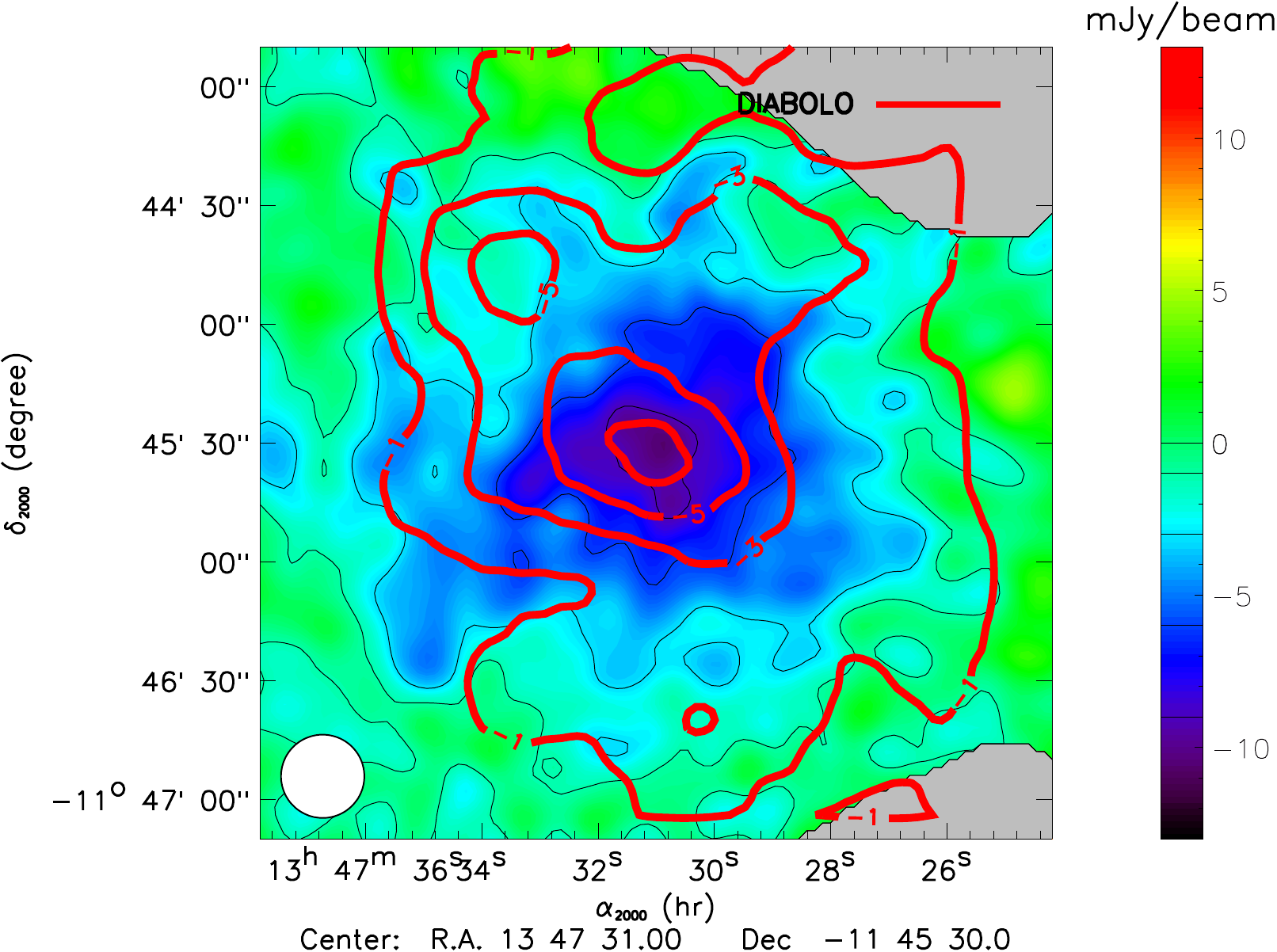}
	\hspace*{0.5cm}
	\includegraphics[width=0.45\textwidth]{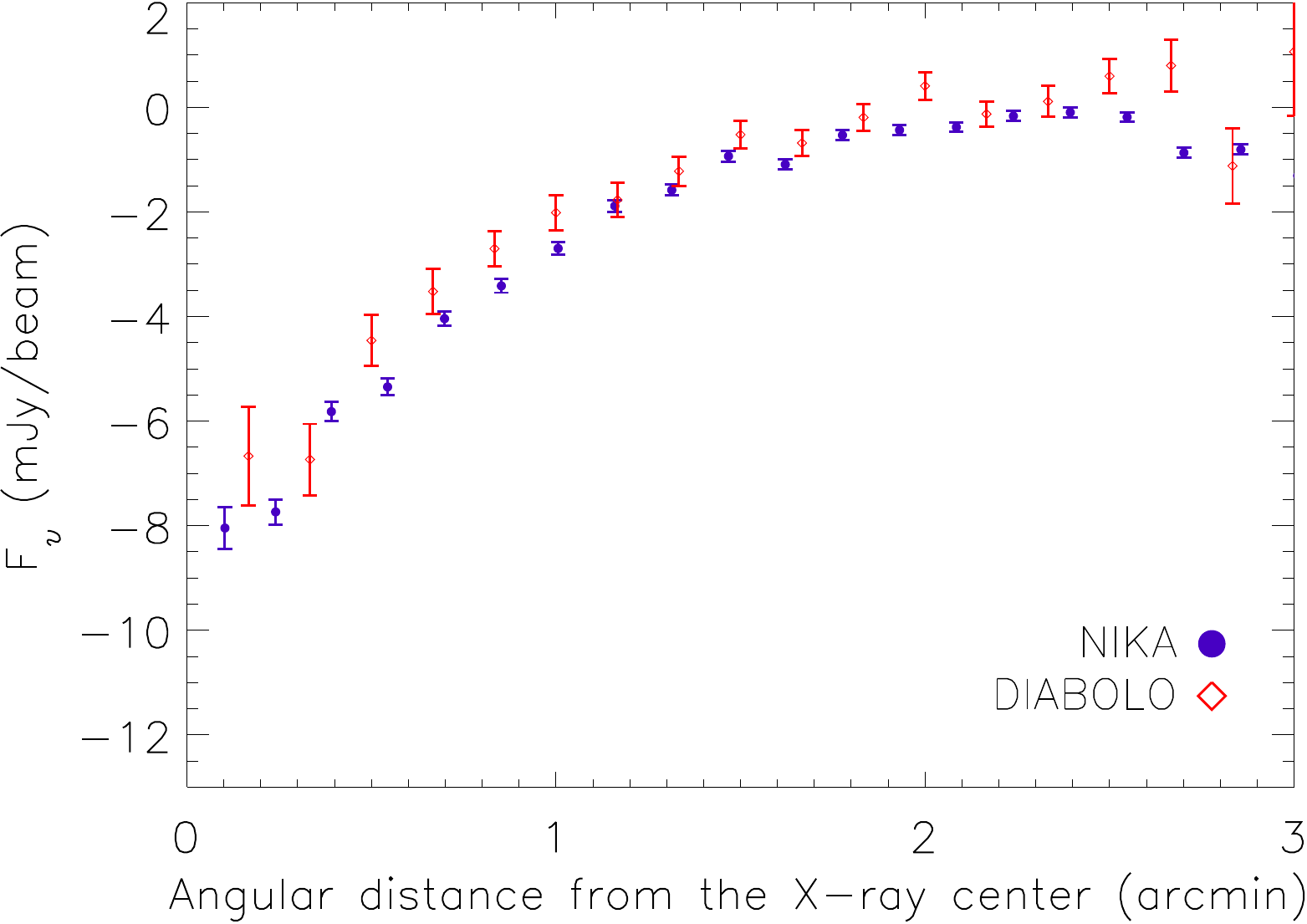}
	\caption{Comparison of \mbox{RX~J1347.5-1145} tSZ maps by {\it DIABOLO} and {\it NIKA} in mJy/beam. Left: {\it NIKA} tSZ map with {\it DIABOLO} contours in red at -1, -3, -5, and -7 mJy/beam. Right: Flux radial profile as measured by {\it NIKA} (purple dots) and {\it DIABOLO} (red diamonds).}
        \label{fig:NIKAvsDIABOLO}
	\end{figure*}

In Fig.~\ref{fig:NIKAvsDIABOLO}, we present the comparison between {\it DIABOLO} and the {\it NIKA} results on \mbox{RX~J1347.5-1145}. {\it DIABOLO} \citep{pointecouteau_1999, pointecouteau_2001} was a bolometric camera that observed \mbox{RX~J1347.5-1145} at the IRAM 30-meter telescope using a dual-band instrument at frequencies corresponding to the {\it NIKA} bands: 140 and 250~GHz. The resolution of {\it DIABOLO} was 22~arcsec at 140~GHz. The data reduction and the instrumental similarities with {\it NIKA} make it a first choice for a direct comparison.

The left panel shows the tSZ {\it NIKA} map with {\it DIABOLO} contours overplotted in red with levels of -1, -3, -5, and -7 mJy/beam (radio source not subtracted in both maps).  We can see that the tSZ maxima and the external part of the cluster match within error bars. The overall amplitude of the signal is slightly higher for {\it NIKA} data than for {\it DIABOLO}. However, this difference is not significant once we account for the  systematic uncertainties given in Table~\ref{tab:table_err}. The right panel of Fig.~\ref{fig:NIKAvsDIABOLO} compares the cluster pressure profile (radio source subtracted and X-rays centered) measured with both instruments. The two profiles are compatible within error bars over the whole radial range, even though {\it NIKA} seems to detect more signal in the inner part of the cluster. The reduced $\chi^2$ associated to the profile difference, which is computed up to a radius of 2.5~arcmin, is equal to 2.35. However, since this does not account for calibration uncertainties, we also give the reduced $\chi^2$ after cross calibrating the two profiles: we obtain $\chi^2 = 1.32$ with a cross-calibration factor of 1.09, which is compatible with our calibration error estimate. In both cases, the tSZ maximum is not located at the \mbox{X-ray} center, in contrast to interferometric {\it CARMA} measurement, but agrees with other single-dish observations.

\subsection{Comparison to {\it XMM} and {\it Chandra} \mbox{X-ray} observations}
\label{sec:comp_x}
\begin{figure*}
	\centering
	\includegraphics[width=0.45\textwidth]{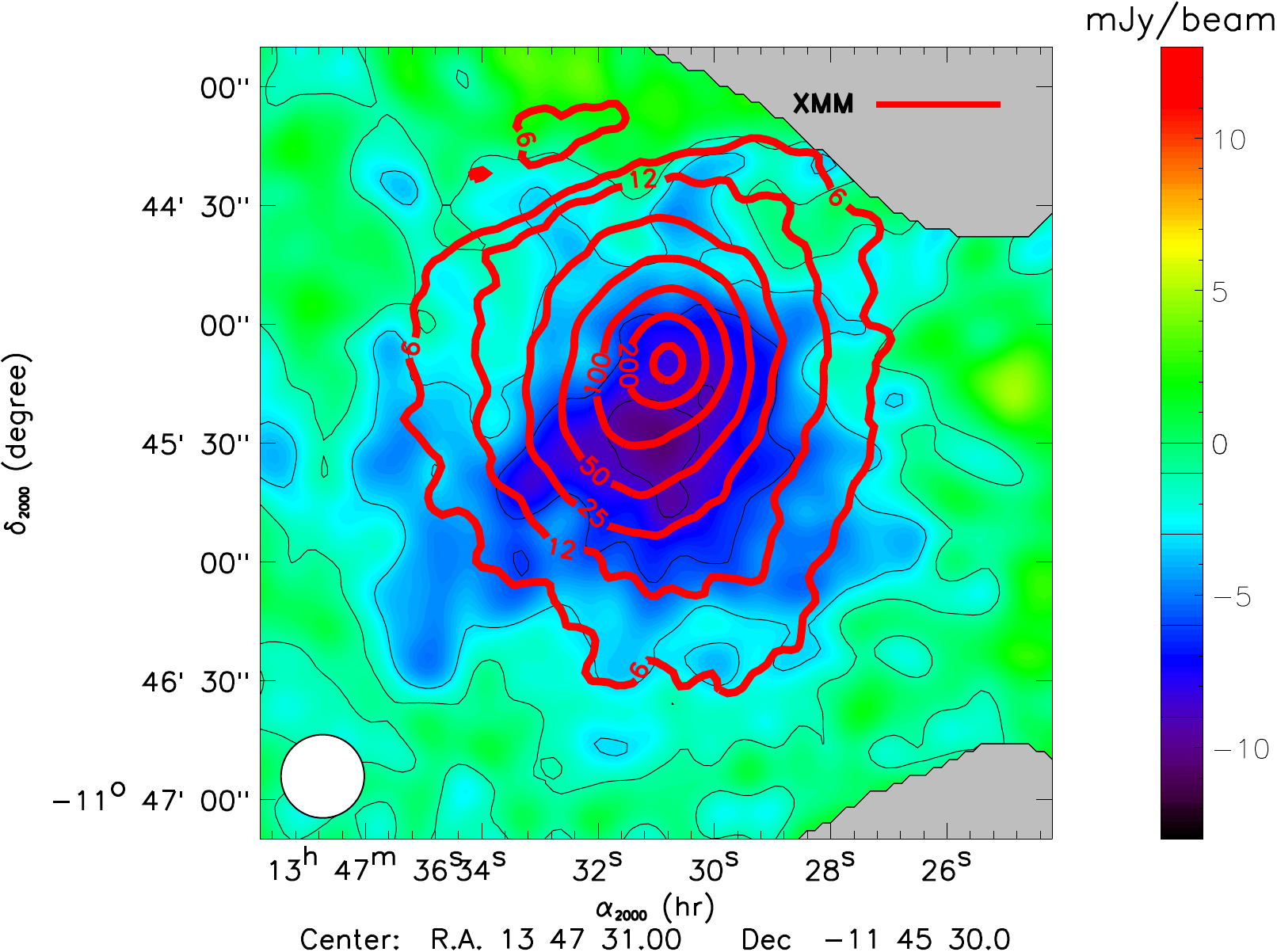}
	\hspace*{0.5cm}
	\includegraphics[width=0.45\textwidth]{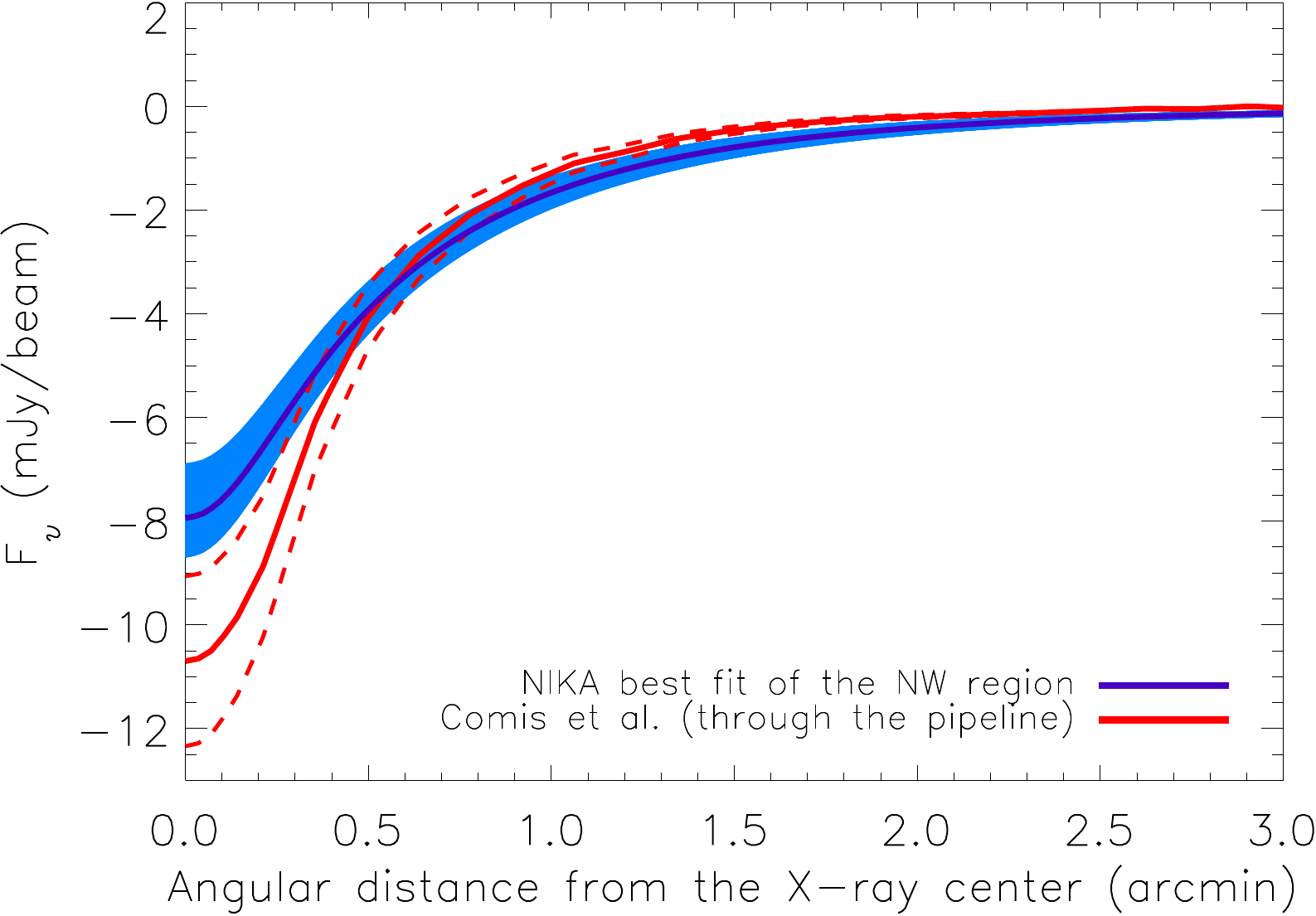}
	\caption{Left: Comparison between the \mbox{RX~J1347.5-1145} {\it NIKA} tSZ map and the {\it XMM} \mbox{X-ray} data \citep[see the work of ][]{gitti_2004,gitti_2005,gitti_2007}. The {\it XMM} map has been smoothed with a 5~arcsec Gaussian filter. The red \mbox{X-ray} contours are in photon counts and correspond to 6 12, 25, 50, 100, 200, and 400. Right: Comparison between the best-fit radial tSZ profile of the {\it NIKA} map, excluding the shock area (see Sect.~\ref{sec:mcmc}) and the profile derived from {\it Chandra}'s data by~\cite{Barbara}, which is processed through the {\it NIKA} data reduction pipeline, as discussed in Sect.~\ref{sec:TOI_ana}. The {\it NIKA} best-fit profile is given in purple with an associated 1$\sigma$ statistical error range filled in blue, and the \mbox{X-ray} model is given in red with the 1$\sigma$ statistical error limit, as two dashed lines. See text for more details on the errors limits.}
        \label{fig:NIKAvsXray}
	\end{figure*}

The {\it XMM} data \citep{gitti_2004} have been used to compute a photon count (exposure corrected) map of \mbox{RX~J1347.5-1145} that we compare to the {\it NIKA} tSZ observations. As seen in Fig.~\ref{fig:NIKAvsXray}, the tSZ peak does not coincide with the \mbox{X-ray} center. The object \mbox{RX~J1347.5-1145} gives a striking example of the power of tSZ data and how it complements \mbox{X-ray}. Moreover, the mismatch between the tSZ and \mbox{X-ray} center gives valuable information on the gas physics at play in the ICM.

The higher resolution of the {\it Chandra} \mbox{X-ray} data has been used for cluster simulation purposes. In particular, the work of \cite{Barbara} uses the publicly available, \mbox{X-ray} derived, pressure profiles of the ACCEPT (Archive of {\it Chandra} Cluster Entropy Profile Tables) clusters \citep{Cavagnolo2009} to constrain the $P_0 $, $r_{\mathrm{s}}$, and $\gamma$ parameters of the gNFW pressure profile (Eq.~\ref{eq:gNFW}). The best-fitting values for \mbox{RX~J1347.5-1145} (Table~\ref{tab:table_clusters}) have been used in the present work to simulate the expected tSZ signal, as explained in Sect.~\ref{sec:sz_simu}. Once processed through the pipeline, the expected profile is compared to the {\it NIKA} best-fit profile, excluding the shocked area (see Sect.~\ref{sec:mcmc}) on the right panel of Fig.~\ref{fig:NIKAvsXray}. The {\it NIKA} best-fit profile and the \mbox{X-ray} model are both given with the 1$\sigma$ error envelope. The error on the {\it NIKA} profile only accounts for statistical uncertainties by sampling the 1$\sigma$ contour of the likelihood of Fig.~\ref{fig:likelihood}. The systematic errors (see Eq.~\ref{eq:best_fit_nika}) are not shown and would result in an overall multiplicative factor on the amplitude ($P_0$) and on the angular scale ($\theta_s$). We choose to include only the error on the parameter $P_0$ (Table~\ref{tab:table_clusters}) for the \mbox{X-ray} model, since it is highly degenerated with $\theta_s$ and $\gamma$. In addition to the \mbox{X-ray} systematic uncertainties, the associated systematic error, which is not included in Fig.~\ref{fig:NIKAvsXray}, arises mainly from the unit conversion coefficient (Jansky per beam to Compton parameter: $y=10^{-3} \equiv 11.8 \pm 1.2$ mJy/beam). It has a similar effect as the error on the $P_0$ {\it NIKA} best-fit value. The two profiles agrees within systematic and statistical uncertainties. 

	\begin{table}
	\begin{center}
	\begin{tabular}{llccc}
	\hline
	\hline
	 $P_0 \ (\mathrm{keV/cm}^3)$ & $3.29 \pm 0.50$ \\
	 ($\alpha$, $\beta$, $\gamma$) & ($0.9$, $5.0$, $0.00 \pm 0.05$) \\
	 $r_{\mathrm{s}}$ (kpc) & $406 \pm 23$ \\
	 $\theta_{\mathrm{s}}$ (arcsec) &  $70 \pm 4$\\
	\hline
	\end{tabular}
	\end{center}
	\caption{Modeling of the pressure profile of \mbox{RX~J1347.5-1145} using the fit of {\it Chandra} data \citep{Barbara}. We note that $\alpha$ and $\beta$ have been fixed to the best-fitting values, as obtained by \cite{nagai_2007}, see \cite{mroczkowski_2009} for an errata.}
	\label{tab:table_clusters}
	\end{table}

\subsection{Comparison to other high resolution tSZ data}
\label{sec:comp_otherSZ}
We also compare the {\it NIKA} data with state-of-the-art sub-arcmin resolution data: {\it MUSTANG} \citep{mason_2010, korngut_2011} and {\it CARMA} \citep{plagge_2012} observations. Since these two instruments are in many ways different from {\it NIKA}, we limit ourselves to a qualitative comparison.

The instrument {\it MUSTANG} uses the single dish 100-meter Green Bank Telescope in Virginia, USA. It operates at 90~GHz with a 8~arcsec resolution. At 90~GHz, the central radio source of \mbox{RX~J1347.5-1145} is very bright compared to the tSZ decrement.  In the case of  {\it MUSTANG}, the removal of the atmospheric noise filters angular scales that are larger than about 60~arcsec. In that sense, the {\it NIKA} and {\it MUSTANG} are complementary. The instrument {\it MUSTANG} is able to measure the structural property of \mbox{RX~J1347.5-1145} at scales ranging from $\sim 10 - 60$~arcsec, while the {\it NIKA} map is reliable in the range of $\sim 20 - 200$~arcsec. The two instruments agree on the morphology of \mbox{RX~J1347.5-1145} at intermediate scales (the inner part of the -6 mJy contour on Fig.~\ref{fig:rxj}). The tSZ maximum coincides and the overall distribution of the tSZ signal is consistent on both observations. The excess seen in the region 2 of Fig.~5 in~\cite{mason_2010} does not show up clearly in the {\it NIKA} map. However, the spatial scales of this feature are smaller than 10~arcsec, and it is likely smoothed out by the {\it NIKA} beam.

The instrument {\it CARMA} is a multifrequency interferometer \citep{plagge_2012}. For \mbox{RX~J1347.5-1145} observations, they were made of 23 antennae of 3.5, 6.1, and 10.4 meter operating in three configurations at 31~GHz, 86~GHz, and 90~GHz for a total of 41.7 hours of unflagged on-source observation. Due to the complexity of combining the data in different configurations, the {\it CARMA} transfer function is not simple. Nevertheless, {\it CARMA} and {\it NIKA} agree well on scales greater than about 30~arcsec. At smaller scales, {\it CARMA} and {\it NIKA} disagree on the position of the tSZ peak.

\subsection{Point source contamination effects}
\label{sec:ps_sub_effect}
As mentioned above, the effect of point source contamination is an issue in single-dish observations. In this work, the radio source located near the \mbox{X-ray} center (within 3~arcsec) can affect our results. The {\it CARMA} data suggest that the flux of the source is underestimated when removed from single-dish data. Therefore, we have reprocessed the {\it NIKA} data by assuming that the source was 3$\sigma$ brighter than its nominal value ({\it i.e.}, 5.3 mJy instead of 4.4 mJy at 140~GHz). We obtain that the tSZ maximum is still located at the shock position. The difference in tSZ amplitude between the \mbox{X-ray} center and the shock is reduced but still inconsistent with {\it CARMA}, even though the discrepancy is smaller. Moving the tSZ maximum to the \mbox{X-ray} center would require a 5$\sigma$ positive shift of the flux of the radio source. The best-fit pressure profile parameters, $P_0$ and $r_{\mathrm{s}}$, are affected by less than 1$\sigma$ (statistical only) by the point source subtraction. This is consistent with what we observed in simulations (see Sect.~\ref{sec:valid_pipe_simu}). For the infrared source Z2, 20~arcsec NE from the \mbox{X-ray} center, we have tested adding a 0.64~mJy source at its position, which is the upper limit that we have estimated in Sect.~\ref{sec:2band_decor}. The changes in our results are negligible for the location of the tSZ maximum and for the $r_{\mathrm{s}}$ and $P_0$ values; they change by less than 0.6$\sigma$ (statistical only). The Z1 infrared source is located in the external part of the cluster and does not affect any of our results.

%% file: 08_conclusion.tex
The cluster RX~J1347.5-1145 is an ongoing merger, among the most-studied galaxy clusters at arcmin angular scales, making it a good target for the first tSZ observations with the {\it NIKA} prototype camera. Using a dual-band decorrelation with a high resolution instrument, we have imaged the tSZ morphology of the cluster from the core to its outer region. The detailed data analysis is specific to KIDs and to tSZ observations and has been validated on simulations. The observation of \mbox{RX~J1347.5-1145} constitutes the first tSZ observations with an instrument based on KIDs.

The reconstructed tSZ map of \mbox{RX~J1347.5-1145} is reliable on scales going from about 20 to 200~arcsec and shows a strong southeast extension that corresponds to the merger shock, as expected from the overpressure caused by the ongoing merger. We detect the non-alignment of the tSZ maximum and the \mbox{X-ray} center, which agrees with other single-dish data but disagrees with {\it CARMA} interferometric data. The tSZ extension is also observed in the radial flux profile of the cluster and the residual of the map with respect to the modeling of the relaxed part of the cluster. The generalized NFW fit of the NW region enables us to constrain the cluster pressure profile parameters $\theta_s$ and $P_0$. The pressure profile derived from \mbox{X-ray} agrees with this tSZ best-fit model.

The tSZ map and the radial profile measured with {\it NIKA} have been compared to {\it DIABOLO} observations at the same telescope with similar resolution and frequency coverage. The agreement between the two maps validates the tSZ observations presented in this work. In addition, the {\it NIKA} prototype map agrees with state-of-the-art sub-arcmin resolution tSZ observations, {\it MUSTANG} (90~GHz and 8~arcsec resolution) and {\it CARMA} (30 -- 90~GHz and $\sim$ 15~arcsec resolution) except for the tSZ peak position. The comparison shows that it is complementary to these experiments.

In this paper, KID arrays of the {\it NIKA} prototype have been proven to be competitive detectors for millimeter wave astronomy and in particular for the observation of galaxy clusters via the tSZ effect. The next generation instrument, {\it NIKA2}, consists of about 1000 detectors at 140~GHz and 4000 at 240~GHz with a field of view of  $\sim 6.5$ arcmin. With these characteristics, {\it NIKA2} is able to provide large high-resolution mapping of clusterss making it an ideal instrument for high-resolution observations of intermediate to large distance clusters of galaxies. The instrument {\it NIKA2} will be well adapted for a follow-up of unresolved sources in the {\it Planck} cluster sample \citep{Planck_survey}. 